\documentclass[12pt, draftclsnofoot, onecolumn]{IEEEtran}
\usepackage{graphicx} 
\usepackage{amsmath,bm,mathtools} 
\usepackage{amssymb}
\usepackage{cite}
\usepackage{colortbl}
\usepackage{subcaption}
\usepackage[shortlabels]{enumitem}
\usepackage{float}
\usepackage{breqn}
\usepackage{lipsum,graphicx}
\usepackage{cite}
\usepackage{algorithm}
\usepackage{algorithmic}
\usepackage{lipsum}
\usepackage{mathtools}
\usepackage{cuted}
\usepackage{stfloats}
\usepackage{siunitx}

\usepackage{scalerel,stackengine}
\usepackage{fontenc}
\usepackage{multicol}
\usepackage{multirow}
\usepackage{xcolor}
\usepackage{diagbox}
\usepackage{soul}
\usepackage{xcolor}
\usepackage{epstopdf}

\begin{document}
 \title{OTFS Channel Estimation and Detection for  Channels with Very Large Delay Spread}
	\author{Preety Priya,~\IEEEmembership{Member,~IEEE}, Yi Hong,~\IEEEmembership{Senior Member,~IEEE}, and \\Emanuele Viterbo, ~\IEEEmembership{Fellow,~IEEE}
 \thanks{Preety Priya is with the Department of Electronics and Communication Engineering, National Institute of Technology Calicut,
Kozhikode 673601, India (e-mail: preetypriya@nitc.ac.in).}
 \thanks{Yi Hong and Emanuele Viterbo are with the Department of Electrical and Computer Systems Engineering, Monash University at Clayton, Clayton, VIC 3800,
Australia (e-mail: yi.hong@monash.edu, emanuele.viterbo@monash.edu).}
		\thanks{This research work is supported by the Australian Research Council (ARC) through the Discovery project: DP200100096. }	
	}
\markboth{IEEE Transactions on Wireless Communications,~Vol.~XX, No.~XX, XXX~2024}%
{Shell \MakeLowercase{\textit{et al.}}: A Sample Article Using IEEEtran.cls for IEEE Journals}

{}
\maketitle

\begin{abstract}
   In low latency applications and in general, for {\em overspread} channels, channel delay spread is a large percentage of the transmission frame duration.
  In this paper, we consider OTFS in an {\em overspread} channel exhibiting a delay spread that exceeds the block duration in a frame, where traditional channel estimation (CE) fails.
  We propose a two-stage CE method based on a delay-Doppler (DD) training frame, consisting of a dual chirp converted from time domain and a higher power pilot.
  The first stage employs a DD domain embedded pilot CE to estimate the aliased delays (due to modulo operation) and Doppler shifts, followed by identifying all the underspread paths not coinciding with any overspread path. The second stage utilizes time domain dual chirp correlation to estimate the actual delays and Doppler shifts of the remaining paths. This stage also resolves ambiguity in estimating delays and Doppler shifts for paths sharing same aliased delay. 
  Furthermore, we present a modified low-complexity maximum ratio combining (MRC) detection algorithm for OTFS in overspread channels. Finally, we evaluate performance of OTFS using the proposed CE and the modified MRC detection in terms of normalized mean square error (NMSE) and bit error rate (BER).
  \end{abstract}
\IEEEpeerreviewmaketitle
\section{Introduction}
 In the realm of next-generation wireless communications with high mobility applications, orthogonal time frequency space (OTFS) modulation exhibits great potential in combating severe Doppler effects. OTFS  multiplexes information symbols in the delay-Doppler (DD) domain \cite{hadani2017orthogonal,delay_Doppler_book}, where a time-varying channel can be represented as a sparse 2D quasi-time invariant channel. This simplifies channel estimation (CE) and facilitates the equalization process of the time-varying channel. 
 
 Prior work related to OTFS CE can be found in  \cite{delay_Doppler_book,VTC_ChaEst_2018,embedded_pilot_raviteja,Shen_OMP_2019,Chocks_multi_user_MIMO_ch_est_2020,Ravi_Radar,Liu_uplink_MIMO_ch_est_2020,Zhao_SBL_2020,Shi_TWC_OTFS,Suraj_BSBL_2021, learning_choks,naikoti2021signal_choks,Mishra_superimposed_pilots_ch_est_2022, jesbin_superimposed,tharaj_OTSM,Thomas_ChaEst} and references therein.
 Specifically, different CE strategies have been proposed including the use of embedded pilot(s) \cite{VTC_ChaEst_2018,embedded_pilot_raviteja,Chocks_multi_user_MIMO_ch_est_2020,Ravi_Radar,Zhao_SBL_2020,Suraj_BSBL_2021,tharaj_OTSM,Thomas_ChaEst}, superimposed pilots \cite{Mishra_superimposed_pilots_ch_est_2022, jesbin_superimposed}, interleaved pilots \cite{learning_choks}, and specific training sequences \cite{Shen_OMP_2019,Liu_uplink_MIMO_ch_est_2020,Shi_TWC_OTFS}.
 
 Regarding OTFS detection, prior work has been developed in \cite{Li_Detection,Zemen_Detection,raviteja2018interference,Tiwari_Detection,surabhi2019low_comp_lin_equal,pandey2021low,Zou_2021_ICC,ge2021receiveroversampling,Hanzo_TVT,Zhang_2021_AMP,Li_2021_EP,Yuan_2022_UAMP,tharaj_rake_MRC_conf,tharaj_rake_MRC_journal,tharaj_universal_MRC,PP_Oversam_MRC,DNN_Cholk,CNN_Enku,Calderbank_learning,naikoti2021signal_choks,Fan_survey} and references therein.
 Various detection methods were proposed including zero-forcing (ZF) and minimum mean square error (MMSE) detection \cite{Zemen_Detection,Tiwari_Detection,surabhi2019low_comp_lin_equal, pandey2021low,Zou_2021_ICC,naikoti2021signal_choks,Fan_survey}, decision feedback equalization
(DFE) \cite{hadani2017orthogonal,Li_Detection}, message passing algorithm (MPA) and its variants \cite{raviteja2018interference,ge2021receiveroversampling,Hanzo_TVT,Zhang_2021_AMP,Li_2021_EP,Yuan_2022_UAMP}, neural network based detection \cite{DNN_Cholk,CNN_Enku,Calderbank_learning}, and maximum ratio combining (MRC) detection \cite{tharaj_rake_MRC_conf,tharaj_rake_MRC_journal,tharaj_universal_MRC,PP_Oversam_MRC}. It was noted in  \cite{Fan_survey} that MPA and its
variants offer excellent performance but with relatively 
high complexity increasing with modulation order. 
In contrast, the MRC detection can offer a performance comparable to standard MPA with a very low complexity independent of modulation order. 
Hence, in this paper, we focus on the low-complexity {\em MRC detection}.

Overall, prior CE schemes and detection methods were tailored for {\em underspread channels}, where delay spread does not exceed the first block of the OTFS frame.   However, a channel can be overspread, particularly in low-latency high mobility communications, where the system has a fixed bandwidth  and a very limited frame size. Increasing the duration of a block (or decreasing subcarrier spacing) to meet the underspread condition on maximum delay spread of the channel would require to increase the number of subcarriers. As a consequence, the number of blocks must decrease to satisfy the low latency constraint on the fixed frame duration. However, decreasing such a number reduces the maximum allowable channel Doppler spread.

Typically, an {\em overspread channel} exhibits a very long delay spread exceeding the block duration in a frame, resulting in {\em aliased delays} in the DD domain due to OTFS modulo operation. Under such settings, traditional CE fails, since it estimates only the aliased delays, but not the {\em actual delays}. Further, the multiple paths sharing same aliased delay can incur ambiguity in estimating actual delays and Doppler shifts.
In addition, the low complexity MRC detection method developed for {\em underspread channels}, needs to be modified for {\em overspread channels}.  
In this paper, we first propose a two-stage CE scheme for OTFS  in such overspread channels. The proposed CE uses a DD training frame consisting of a dual chirp converted from time domain and a higher power pilot. This was inspired by \cite{dual_chirp}, where a dual chirp sequence was used for time synchronization in underspread channels in conjunction with the Radon-Wigner transform.

In the first stage of the proposed CE, a DD domain embedded pilot is used to estimate
aliased delays and Doppler shifts,
followed by identifying all the underspread paths not coinciding
with any overspread path. The second stage utilizes time domain
dual chirp correlation to estimate the actual delays and Doppler
shifts of the remaining paths. In some rare cases, refinement steps can be used to resolve
ambiguity in estimating delays and rectify incorrect estimates of Doppler shifts for paths
sharing same aliased delay. 

Furthermore, we present a modified low-complexity MRC detection algorithm for reduced zero-padding OTFS (RZP-OTFS) in overspread channels. Complexity of the proposed CE and the modified MRC detection is discussed.
 Finally, we evaluate the performance of the proposed CE and MRC detection in terms of normalized mean square error (NMSE) and bit error rate (BER). 

The rest of the paper is organized as follows. Section \ref{Sec:System}
presents system model. In Section \ref{Sec:CE}, a two-stage CE scheme is proposed for overspread channels, followed by complexity discussions. Section \ref{Sec:Detection} presents a modified low-complexity MRC detection algorithm, and discusses its complexity and convergence. Section~\ref{Sec:Simulation} provides simulation results. Finally,  
conclusions are drawn in Section~\ref{Sec:Concl}.

\indent \textbf{Notations:}  $x, \textbf{x}, \textbf{X}$ represent scalar, vector, and matrix, respectively; $(\cdot)^H$, $(\cdot)^T$, $(\cdot)^*$, $\lvert \cdot \rvert$, $\lfloor\cdot\rfloor$, $[\cdot]_M$ denote the Hermitian, transpose, conjugate, modulus, nearest integer, modulo $M$, respectively; $\textbf X[m,n]$ represents element in $m^{th}$ row and $n^{th}$ column of matrix $\textbf X$;  and $\textbf x[n]$ denotes $n^{th}$ element of vector $\textbf x$; $\text{vec}(\textbf A)$ denotes the column vectorization of matrix $\textbf A$; $\text{vec}^{-1}(\textbf a)$ denotes the inverse operation of vectorization; $\mathcal B$ represents a set and $|\mathcal B|$ is the cardinality of set $\mathcal B$; $\jmath$ represents unit imaginary number, defined as $\sqrt{-1}$;
$\textbf F_M$ is the $M$-point normalized discrete Fourier transform (DFT) matrix with its entries as $\textbf F_M[m,n]=\frac{1}{\sqrt M}e^{-\jmath2\pi  m n/M}$. The operator ${\bf a}\circ {\bf b}$ represents element-wise multiplication of two same-size vectors ${\bf a}, {\bf b}$.

\section{System Model}\label{Sec:System}
	We consider an OTFS system operating on a multipath ($L$ paths) time-varying wireless
channel with maximum delay spread $\tau_{\max}$ and maximum
Doppler shift $\nu_{\max}$. In this system, we assume that each frame has a duration $T_f=NT$, occupying a bandwidth $B=f_s = 1/T_s=M\Delta f$, where $M$ is the number of subcarriers spaced by $\Delta f = 1/T$, $N$ is the number of blocks per frame,  $f_s$ represents the Nyquist sampling rate, and $T_s$ is the Nyquist sampling interval. This system can support the transmission of $MN$ $Q$-QAM information symbols each with energy $E_s$.
 
Further, we consider that the OTFS system adopts an inverse discrete Zak transform (IDZT)  \cite{Saif_IDZT} converting the QAM information symbols matrix ${\bf X}\in{\mathbb{C}^{M\times N}}$ in the DD domain to the time-domain signal sample vector as \cite{delay_Doppler_book}
 \[{\bf s} = \text{vec}(\widetilde{\textbf X})= \text{vec}(\textbf X \textbf F_N^H )\in{\mathbb{C}^{MN\times 1}}, 
 \]
 where
\begin{equation}\label{EQ:IDZT}
    \widetilde{\textbf X}=[\tilde{\bf x}_0^T,\ldots,\tilde{\bf x}_{M-1}^T]^T= \textbf X \textbf F_N^H \in{\mathbb{C}}^{M\times N}
\end{equation} 
is the delay-time (DT) samples matrix, where $\tilde{\bf x}_m\in{\mathbb{C}}^{N\times 1}$, $m\in[0, M-1]$, are DT domain OTFS symbols.
 For the $i$-th path of the channel, where $i \in [0,L-1]$, 
 the complex path gain is $h_i$, and
the actual delay and Doppler shift are $\tau_i = \frac{l_i}{M\Delta f}$ and $\nu_i = \frac{k_i}{NT}$, where $l_i$ and $k_i$ are the {\em normalized delay} and {\em normalized Doppler shift}, respectively. Note that  $l_{\max}=M\Delta f\tau_{\max}$ denotes the {\em normalized delay} associated with $\tau_{\max}$. 

For sufficiently high delay and Doppler resolutions, we assume $l_i$ and $k_i$ are {\em integers}. We also assume that {\em each delay is associated with only one Doppler}.

 Different from the standard OTFS settings, in this paper, we assume that the channel is {\em overspread}, i.e., $\tau_{\max}\nu_{\max}>1$. Under such an assumption, we have some $l_i \ge M$ and the normalized
Doppler shifts\footnote{ Note that the joint overspreading in delay ($\tau_{\max}>T$) and in Doppler ($|\nu_{\max}|>\frac{\Delta{f}}{2}$) would occur for speeds beyond 30,000 km/h, for standard values of $M$ and $N$. Dealing with overspread Doppler would require a distinct approach, which falls outside of the scope of the current paper and will be considered in our future work.} $-\frac{N}{2} < k_i \le \frac{N}{2}$.  


 
 \section{Proposed Channel Estimation for overspread channels}\label{Sec:CE}
For OTFS in an {\em overspread channel}, we propose  a novel CE scheme to estimate path gain, normalized delay, and normalized Doppler shift of each path.  
 The scheme relies on the following training frame, and operates in a two-stage process.
 \subsection{Proposed Training Frame Arrangement}
  
Let us define a {\em generic dual-chirp pulse} of duration $M$ samples, delayed by $l$ and Doppler shifted by $k$ as
\begin{equation}\label{EQ:generic_Chirp}
    {\textbf p}[l,k,q]\triangleq
    \begin{cases}A\sum\limits_{a\in \pm 1} e^{\jmath2\pi\Big\{\Big(\frac{f_oT}{M}+\frac{k}{MN}\Big)(q-l)+\frac{a}{4M}(q-l)^2\Big\}}\\ 
    \hspace{2.2cm} \text{for~} q\in[l,l+M-1]\\[1ex] 0  \hspace{2cm}\text{otherwise},\end{cases}
\end{equation}
where $A$ is the amplitude and $f_o$ is the center frequency of the dual chirp formed by the sum of two chirps with opposite chirp rates $\pm \frac{B}{2T}$. 

Let ${\textbf p}[0,0,q]$ for $q\in [0,MN-1]$ be the time domain {\em transmit dual chirp} and $\textbf X_{\rm c}=\text{DZT}({\textbf p}[0,0,q])$ be its corresponding DD domain representation. We arrange a single pilot $x_{\rm p}$ and $\textbf X_{\rm c}$ in the DD domain to form the training frame as
\begin{equation}\label{EQ:TrainFrame}
    \textbf X_{\rm t}[l,k]=\begin{cases}\textbf X_{\rm c}[0,0]+x_{\rm p}&~~l=k=0 \\[1ex] \textbf X_{\rm c}[l,k]&~~l\neq 0,k\neq 0, \end{cases}
\end{equation}
for $l\in [0,M-1], k\in [0,N-1]$, which leads to the time-domain training signal vector  
 \begin{equation}\label{td_OTFS}
    \textbf s_{\rm t}\triangleq \text{IDZT}(\textbf X_{\rm t})=\text{vec}\Big(\textbf X_{\rm t} \textbf F_N^H\Big)\in{\mathbb{C}^{MN\times 1}}. 
\end{equation}
Following the input-output relation of OTFS system  in \cite{delay_Doppler_book}, the time-domain received training vector can be expressed 
as 
\begin{equation}\label{td_rx_OTFS}
\textbf r_{\rm t}[q]=\sum_{i=0}^{L-1}h_ie^{\jmath\frac{2\pi}{MN}k_i(q-l_i)} \textbf s_{\rm t}[q-l_i]+\textbf w[q],
\end{equation}
 where $\textbf w[q]$ are the additive white Gaussian noise (AWGN) samples with zero mean and variance $\sigma^2_w$. Further, the DD received training matrix $\textbf Y_{\rm t}\in{\mathbb{C}^{M\times N}}$ is obtained as \cite{delay_Doppler_book}
\begin{equation}
    \textbf Y_{\rm t}\triangleq \text{DZT}(\textbf r_{\rm t})=\text {vec}^{-1}(\textbf r_{\rm t})~\textbf F_N.
\end{equation}
 
Assuming only the dual-chirp pulse is transmitted, i.e., $\textbf s_{\rm t}[q] = \textbf p[0,0,q]$, then (\ref{td_rx_OTFS}) becomes 
\begin{equation}\label{discrete_td_rx_OTFS}
     \textbf r_{\rm c}[q]=\sum_{i=0}^{L-1}h_i {\textbf p}[l_i,k_i,q] + \textbf w[q].
\end{equation}
 This received chirp vector $ \textbf r_{\rm c} \in\mathbb{C}^{MN\times 1}$ will be used later for channel estimation.

\begin{figure} 
\begin{center}
\setlength{\arrayrulewidth}{0.3mm}
\setlength{\tabcolsep}{5pt}
\renewcommand{\arraystretch}{1}
\begin{tabular}{|p{3.5mm}||p{7mm}|p{9mm}|p{9mm}|p{7mm}|p{7mm}|p{7mm}|} 
\hline
 $_\ell\backslash^k$&0&1&2&3&4&5 \\
\hline \hline
0& {\sf X} $_{0}$ & & & & &   \\
\hline
1& & &  & & {\sf O} $_{1}$&   \\
 \hline
2& & {\sf X} $_{2}$& & {\sf O} $_{3}$ & &   \\
 \hline
3& & & & & &   \\
 \hline
4& & &{\sf O\hspace{-3mm}X} $_{4,5}$ & & &{\sf O} $_{6}$    \\
 \hline
5& & & & & &   \\
 \hline
6& & {{\sf O\hspace{-4mm}O}} $_{7,8}$ & & & &   \\
 \hline
7& & & & & &    \\
 \hline
\end{tabular}
\end{center}
\caption{DD received pilot echos with $M=8,N=6$ with 9 paths $(l_i,k_i)=\{(0,0),(9,4), (2,1),(10,3),(4,2),(12,2)$, $(20,5),(14,1),(30,1) \}$. The path index  $i=0,\ldots, 8$ is given in each cell,  
{\sf X} represents an underspread path  ($l_i<M$) and {\sf O} an overspread path ($l_i\geq M$). 
} \label{over_ch_dd}
\end{figure}
\subsection{Proposed Two-Stage CE}
Let $\mathcal L=\{l_i\}, i\in [0,L-1]$ denote the set of the normalized delays of the channel paths. We define  
\[
{\ell}_i \triangleq[l_i]_M~~~{\text{for}~~l_i\in{\mathcal L}}
\]
or equivalently
\begin{equation}\label{EQ:Mod_M}
l_i = {\ell}_i +b_i M~~~{\text{for}~~b_i\in[0,N-1]}
\end{equation}
where $b_i=\lfloor \frac{l_{i}}{M}\rfloor\in [0,b_{\max}]$ represents the block index associated with $l_i$, where $b_{\max}=\lfloor \frac{l_{\max}}{M}\rfloor < N$ is the block index corresponding to the maximum delay $l_{\max}$.
 
Let the {\em aliased paths set} be defined as
\begin{equation}
    {\tilde{\mathcal L}}\triangleq\{{\ell}_i|l_i \ge M\},
\end{equation}
 containing only the wrapped delays affected by the modulo $M$ operation.
 
 In the following, we {\em omit the subscripts in channel parameters} for simplicity of the description, unless {\em otherwise needed}.
 
 Fig. \ref{over_ch_dd} shows the DD received pilot echos resulting from an overspread channel, where only the aliased delays can be observed. When multiple paths share the same aliased delay $\ell$ with same Doppler shift, there is ambiguity in determining delays.
 When paths have same aliased delay $\ell$ but with distinct Doppler shifts, there is ambiguity in pairing the correct Doppler shift $k$ to each path with actual delay $l$, as shown in the example below. 

 \textit{Example 1:} In Fig. \ref{over_ch_dd}, paths 4, 5 have same aliased delay $\ell=4$ and same Doppler shift $k= 2$, and paths 7, 8 share same $\ell= 6$ with same Doppler shift $k=1$, resulting in delay ambiguity. Further, paths 2 and 3 share same $\ell$ with distinct Doppler shifts, leading to Doppler ambiguity, and  similarly for paths 4 (or 5) and 6.  Overall, paths 4, 5, and 6 encapsulate both delay and Doppler ambiguity.

Hence, the goal is to estimate the actual delays $l$, Doppler shifts $k$, and gains $h$ for all the $L$ paths, where $l$  depends on both $\ell$ and $b$, and in particular, for underspread paths
\begin{equation}\label{eq:bi_underspred}
  l = \ell  ~~\text{and} ~~b=0  
\end{equation}
and for overspread paths,
\begin{equation}
   l \neq \ell  ~~\text{and} ~~ b\neq 0. 
\end{equation}

 In channel estimation, once we determine the estimates $\hat l =\hat \ell +\hat b M$ and $\hat k$ of a path, the pilot $x_{\rm p}$ can be used to estimate the path gain. 

   \begin{figure} 
	 	\centering		\includegraphics[width=.5\linewidth]{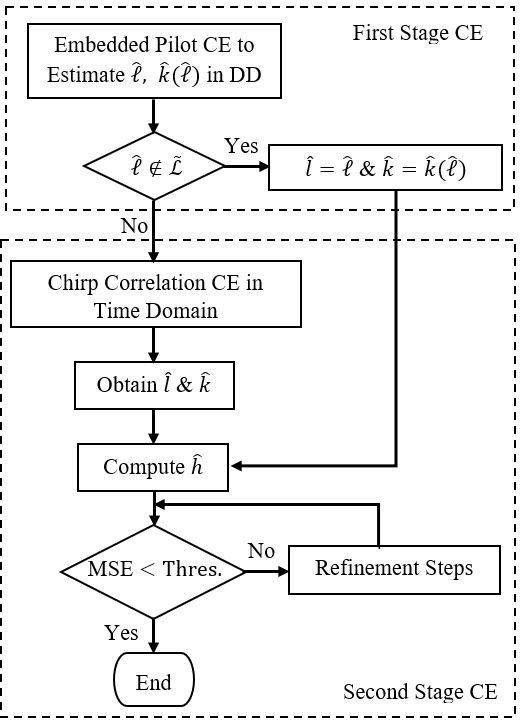}
	 	\caption{\small Flow Chart of Two-Stage CE}
	 	\label{flow_ch_main}
	 \end{figure}

Hence, in order to determine the actual delays, Dopplers shifts of all the paths without ambiguities,
we propose a two-stage channel estimation process (see Fig. \ref{flow_ch_main} and Algorithm \ref{Alg_first_stage}).

 The first stage CE adopts an embedded pilot in DD domain to estimate the aliased delay and Doppler shift, followed by identifying whether it is an underspread path not coinciding with an overspread one (e.g., path 0 in Fig. \ref{over_ch_dd}). 
If this is true then we obtain the estimates of the normalized delay $\hat l=\hat \ell $, the Doppler shift, and $\hat{b}=0$.

Otherwise, for overspread delays  and underspread delays coinciding with overspread ones, we need to proceed to the second stage based on chirps correlation in time-domain (see \textit{ChirpCorr} function in Appendix) to estimate normalized delays and Doppler shifts. 
This chirp correlation method also resolves the Doppler ambiguity which occurs when multiple paths have the same aliased delay but with different Doppler shifts. 

After obtaining the delay-Doppler estimates, we compute the channel coefficients.
Thereafter, the accuracy of the estimation is evaluated by the mean square error (MSE) between the received training frame $\textbf r_{\rm t}$ and its reconstructed counterpart $\hat{\textbf r}_{\rm t}$ using the estimated channel parameters. If the MSE is beyond a threshold, the refinement steps (see \textit{Refine1} and \textit{Refine2} functions in Appendix) will be conducted until it is exhausted or the threshold is met. The refinement steps resolve the ambiguity in determining actual delays with same Doppler shift and rectify any incorrect Doppler estimates from chirp correlation CE, given in the \textit{ChirpCorr} function. 
\subsection{First Stage CE using Embedded Pilot}\label{First_CE}

In the DD domain, we assume ${|x_{\rm p}|^2}\gg \frac{{2A}^2}{N}$. Since the dual chirp signal spreads over the entire DD domain, the impact relative to pilot $x_{\rm p}$ is negligible for sufficiently large $N$. Consequently, we have
\begin{equation}\label{eq:thres_ch_est}
  \textbf Y_{\rm t}[\ell,k]=h e^{-\jmath\frac{2\pi}{N} k b}x_{\rm p} +\epsilon [\ell,k],
\end{equation}
 where $\epsilon [\ell,k]$ represents the contribution of chirp, pilot
spreading along Doppler axis, caused by DFT operation on the
DT domain shifted pilot echos in a channel with overspread
delays, and noise in the $(\ell,k)$-th sample in the DD domain. 

Next, we propose a CE procedure by first selecting the delay indices potentially containing pilot echoes, then estimating the corresponding Doppler shifts via an adaptive threshold-based method. For this, we let
\begin{equation}\label{EQ:avg-Power}
P(\ell)=\frac{1}{N}\sum_{k=0}^{N-1}|\textbf Y_{\rm t}[\ell,k]|^2
\end{equation}
be the {\em average power} at the $\ell$-th delay index. 
The potential presence of pilot echoes can be established when $P(\ell)\geq \delta(2A^2/N + \sigma_w^2)$, for some  $\delta>0$.  Let  $\mathcal A =\{\}$. At such $\ell$-th delay, we append the {\em estimate of the aliased delay} $\hat \ell = \ell$ to set $\mathcal A$, and  estimate the corresponding Doppler shifts using an {\em adaptive threshold} $\mathcal T_{\ell}=\alpha P(\ell)$ for some $\alpha>0$, as
\begin{eqnarray}\label{EQ:Embedded}
   \hat k (\hat \ell)=k,~~~{\text{if}} ~~|\textbf Y_{\rm t}[\ell,k]|^2>\mathcal T_{\ell}
\end{eqnarray}
where $\hat k (\hat \ell)$ denotes the Doppler estimate corresponding to $\hat \ell$.
Here, the {\em adaptive threshold} is made to take into account pilot spreading along Doppler axis. 
This is different from the traditional CE scheme in \cite{embedded_pilot_raviteja} using a constant noise variance based threshold.

Next, we append {\em all the estimates of Doppler shifts corresponding to the same} $\hat{\ell}$ to the set \[
 \mathcal K_{{\hat \ell}}=\{\hat k_{\lambda}({\hat \ell})\},\] where $\lambda$ is the index of the set.  

 When multiple paths coincide at the same delay index but with distinct Dopplers, we have $|\mathcal K_{{\hat\ell}}|>1$, otherwise $|\mathcal K_{{\hat\ell}}|=1$.

Then, to check if $\hat{{\ell}}$ belongs to an underspread path not coinciding with an overspread one (i.e., $|\mathcal K_{{\hat\ell}}|=1$), we compute {\em the average power at ${{\hat\ell}}$-th delay excluding pilot power} as 
\begin{equation}\label{EQ:Avg_Pwer_woPilot}
P'({\hat\ell})=\frac{NP({\hat\ell})-\sum_{\lambda=1}^{|\mathcal K_{{\hat\ell}}|}~\Big|\textbf Y_{\rm t}[\hat{{\ell}},\hat k_{\lambda}({\hat\ell})]\Big|^2}{N-|\mathcal K_{{\hat\ell}}|},
\end{equation}
where $P({\hat\ell})$ is given in (\ref{EQ:avg-Power}).
Due to pilot spreading incurred by overspread delays,  we have $P'({\hat\ell}) \ll P'({\hat\ell}')$, where ${\hat\ell}\notin \tilde{\mathcal L}$ is the aliased delays for an {\em underspread path  not coinciding with any overspread path} and ${\hat\ell}^{\prime}\in \tilde{\mathcal L}$ is the aliased delay for an {\em overspread path}. Hence, we determine the underspread path and its channel parameters as 
\begin{equation} \label{undspred_delay_det}
    \hat l= \hat {\ell}, ~~\hat k= \hat k(\hat {\ell}),~~\text{and}~~\hat b=0,~~ \text{if}~~~P'({\hat\ell}) \leq {\mathcal T}^\prime 
\end{equation}
where $\mathcal T^{\prime}=\alpha^\prime \sigma_w^2$,  $\alpha^\prime>0$.  The estimate of the channel gain $\hat h$ of such paths are obtained as \cite{delay_Doppler_book}
\begin{equation}\label{hi_est}
      \hat h=\frac{\textbf Y_{\rm t}[[\hat l]_M, [\hat k]_N]}{x_{\rm p}}.
      \end{equation}

We let $\mathcal H$ be the path list and we append all such paths $\{(\hat l,\hat k, \hat h)\}$ to $\mathcal H$.

For the {\em remaining aliased delays}  with $P'({\hat\ell}') > {\mathcal T}^\prime$, we define a set 
\[{\mathcal J}=\{\hat\ell^{\prime}|P'({\hat\ell}') > {\mathcal T}^\prime\}\]
to be used in the second stage CE. The first stage CE is included in Algorithm \ref{Alg_first_stage}. 

\textit{Example 2:} Consider the example in Fig. \ref{over_ch_dd}. Here, 
$\mathcal K_{0}=\{0\}, \mathcal K_{1}=\{4\}, \mathcal K_{2}=\{1,3\}, \mathcal K_{4}=\{2,5\}, \mathcal K_{6}=\{1\}$, and $\mathcal J=\{1,2,4,6\}$.
 \begin{algorithm}[t]
    		\caption{Two-stage CE }
    		\begin{algorithmic} [1]
     \STATE {\textbf{Input:} $\textbf r_{\rm t}, \textbf s_{\rm t}, \textbf Y_{\rm t}, \gamma, \alpha, \delta,\sigma_w^2, \mathcal T^\prime, \Gamma,\Gamma^\prime, l_{\max}$}
      \STATE{\textbf{Output:} ${\mathcal H} =\{(\hat l_i, \hat k_i, \hat h_i)\}$}
        \STATE {{\bf Initialize:} Path list $\mathcal H=\{\}$, set ${\cal A}=\{\}$, $\mathcal J=\{\}$, $i=1$}

        \STATE{$\%$ ---- Stage 1 CE ----}
        \FOR {$\ell\rightarrow 0~\text{to}~M-1$}
        \STATE{Compute $P(\ell)$ using (\ref{EQ:avg-Power})} 
\IF{$P(\ell)\geq {\delta}(2A^2/N+\sigma_w^2)$}
        \STATE{$\hat \ell \leftarrow \ell$, ${\cal A} \leftarrow  {\cal A} \cup \{\hat\ell\}$, $K_{\hat \ell} =\{\}$   }
        \STATE{
        Estimate all $\hat k (\hat \ell)$ using (\ref{EQ:Embedded}), and  $\mathcal K_{\hat \ell} \leftarrow  \mathcal K_{\hat \ell} \cup \{\hat k (\hat \ell)\}$ }
        \ENDIF
        \ENDFOR
        \STATE{Compute $P'({\hat \ell}), ~{\text{for}~\text{all}} ~\hat\ell \in {\cal A}$ using (\ref{EQ:Avg_Pwer_woPilot})}
        \FOR{$\hat \ell \in{\cal A}$}
        \IF {$P'({\hat\ell}) \leq {\mathcal T}^\prime$}
        \STATE{Estimate $\{(\hat l_i= \hat {\ell},\hat k_i= \hat k_\lambda(\hat {\ell})\in \mathcal K_{\hat \ell},\hat h_i)\}$}
        \STATE{$\mathcal H\leftarrow \mathcal H\cup \{(\hat l_i,\hat k_i, \hat h_i)\}$}
          \STATE{$i\leftarrow i+1$}
        \ELSE
        \STATE{$\mathcal J\leftarrow\mathcal J \cup\{\hat\ell\}$}
        \ENDIF
        \ENDFOR
\STATE{$\%$ ---- Stage 2 CE ----}
      \STATE{$[{\mathcal H}]=\textit{ChirpCorr} (\textbf r_{\rm t}, \textbf s_{\rm t}, \gamma, \sigma_w^2,\Gamma,\Gamma^\prime, l_{\max},i,{\mathcal J}, \mathcal H, $$\mathcal K_{\hat \ell},\forall  \hat\ell\in \mathcal J)$}
      \STATE{Return $\mathcal H$}
    		\end{algorithmic} \label{Alg_first_stage}
    	\end{algorithm}

\subsection{Second Stage CE using Dual Chirps}
To estimate the normalized delay and Doppler shift of the aliased paths in set  $ {\mathcal J}$, we propose a time domain dual chirp correlation method (see \textit{ChirpCorr} function in Appendix). 

We first remove the high power pilot samples from  the received training vector $\textbf r_{\rm t}$ in (\ref{td_rx_OTFS}) to isolate the received dual chirp as
\begin{equation}
    \check{\textbf r}_{\rm c}[q]=\begin{cases}\textbf r_{\rm t}[q] & |\textbf r_{\rm t}[q]|^2\leq\Gamma\\[1ex]0 &  |\textbf r_{\rm t}[q]|^2>\Gamma\end{cases},
\end{equation}
where $\Gamma$ is a threshold dependent on pilot power. 
Then we identify the {\em set of the likely block indices} ${\mathcal B}=\{\hat b_\beta\in [0,b_\max]\}$, corresponding to the potential $\hat l$ using the following steps:
\begin{itemize}
    \item \textbf {Step a:} Compute cross-correlation function between $\check{\bf r}_c$ and the transmitted dual chirp  as
\begin{equation}\label{EQ:Cross_correlation}
R_{\check {\textbf r}_{\rm c},\textbf p}[q] = \sum_{q^\prime =0}^{MN-1-q} \check{\bf r}_{\rm c}[ q+q^\prime]{\bf p}^* [0,0,q^\prime ],
\end{equation}
for $q \in [0, l_{\max}]$.  
   
    \item \textbf {Step b:}  Identify the samples satisfying $|R_{\check {\textbf r}_c,\textbf p}[q]|\geq\Gamma^\prime$, where $\Gamma^\prime$ is a threshold, and collect the corresponding sample indices to form a set $\mathcal Q=\{q~|~~|R_{\check {\textbf r}_{\rm c},\textbf p}[q]|\geq\Gamma^\prime\}$. 
    \item \textbf {Step c:} Find the set $\mathcal B=\{\hat b_\beta\}$ containing the unique values of $\lfloor{\frac{ q}{M}}\rfloor$  for all $q\in \mathcal Q$.
\end{itemize}
We note that the cross-correlation in 
(\ref{EQ:Cross_correlation}) does not sharply peak at the exact delay indices since the received dual chirps are affected by Doppler shifts and do not fully match with the 
${\bf p}^* [0,0,q^\prime ]$
which has zero Doppler shift. 

Next, for a particular aliased delay $\hat \ell\in \mathcal J$, we conduct the following steps to determine all the normalized delays $\hat l$, Doppler shifts $\hat k$, and channel gains $\hat h$ associated with $\hat \ell$. 
\begin{itemize}
    \item \textbf {Step 1:} For all $\beta \in[1,|\mathcal B|]$, obtain all the potential normalized delays as
    \[\hat l(\hat {\ell},\hat b_\beta)=\hat {\ell}+\hat b_\beta M, ~~\forall~\hat b_\beta\in \mathcal B,\]
    and associate each delay $\hat l(\hat {\ell},\hat b_\beta)$ with all the Doppler shifts $\hat k_{\lambda}({\hat\ell})\in\mathcal K_{{\hat\ell}}$, $\lambda \in[1,|\mathcal K_{{\hat\ell}}|]$.
    \item \textbf {Step 2:} Form the corresponding dual chirp  ${\textbf p}[0, \hat k_{\lambda}({\hat\ell}), q]$ 
   and compute its cross-correlation function with  $\check{\bf r}_c$ at $\hat l(\hat {\ell},\hat b_\beta)$ as 
    \begin{equation}\label{EQ:cross-corr}
\textbf{C}[\beta,\lambda]= 
R_{\check{\bf r}_c, {\textbf p}}[\hat l(\hat {\ell},\hat b_\beta)], 
\end{equation}
for all $\beta$'s and $\lambda$'s.
  
 \item  \textbf {Step 3:} For each $\beta\in[1,|{\mathcal B}|]$, find 
 \[(\beta,\lambda^*_\beta)=\arg\max\limits_{\lambda} \textbf C[\beta,\lambda],\]
 to obtain delay-Doppler pair $(\hat l(\hat {\ell},\hat b_\beta), \hat k_{\lambda_{\beta}^*}({\hat\ell}))$. 
\item  \textbf {Step 4:} Take the $|{\mathcal K}_{\hat \ell}|$ largest values from $\textbf C[\beta,\lambda^*_\beta]$  for $\beta\in [1,{|\mathcal B|}]$ with indices set $\mathcal I\subseteq [1,|\mathcal B|]$.  

    \item  \textbf {Step 5:} Determine the delay, Doppler shift, and channel gains estimate set as
\[\mathcal H^\prime=\Big\{\Big(\hat l=\hat l(\hat {\ell},\hat b_{\beta^*}),\hat k=\hat k_{\lambda^*_{\beta^*}}({\hat\ell}), \hat h\Big)\Big|\beta^*\in\mathcal I\Big\}.\]

 Due to the pilot spreading of the overspread delays, the DD domain channel gain estimation in (\ref{hi_est}) cannot be used. Hence, for all such paths, we  estimate $\hat h_i$ for $i$-th path in the time domain as
\begin{equation}\label{hi_est_td}
    \hat h_i=\frac{\textbf r_{\rm t}[\hat l_i]-\sum_{j=0}^{i-1}\hat h_{j}e^{\jmath\frac{2\pi}{MN}\hat k_
    {j}(\hat l_i-\hat l_{j})}\textbf s_{\rm t}[\hat l_i-\hat l_{j}]}{e^{\jmath\frac{2\pi}{MN}\hat k_i}\textbf s_{\rm t}[0]},
\end{equation}
where $\hat h_{j} = 0$ for $j<0$. Note that (\ref{hi_est_td}) is based on canceling the interference from paths with delays less than $\hat l_i$ of $i$-th path.
    \item \textbf {Step 6:} Append $|{\mathcal K}_{\hat \ell}|$ channel path's parameter estimates $\{(\hat l,\hat k,\hat h)\}$ obtained in Step 5 to $\mathcal H$, i.e., $\mathcal H=\mathcal H \cup \mathcal H^\prime$.
    \item \textbf {Step 7:} Compute ${\rm MSE}=\mathcal E\{\textbf r_{\rm t}, \hat{\textbf r}_{\rm t}\}$ as
    \[\mathcal E\{\textbf r_{\rm t}, \hat{\textbf r}_{\rm t}\}=\frac{||\textbf r_{\rm t}-\hat {\textbf r}_t||^2}{MN},\]
    where $\hat{\textbf r}_{\text{t}}$ is the reconstructed training frame using the estimated channel parameters in $\mathcal H$. If  ${\rm MSE}\geq \gamma \sigma_w^2$, where  $\gamma>0$, then potential errors are still present in the channel estimation. This requires the refinement steps as discussed below  until ${\rm MSE}< \gamma \sigma_w^2$ or all the refinement steps are exhausted. 
\end{itemize}
\textit{Example 3:}
In Fig. \ref{over_ch_dd}, we have $\mathcal B=\{0,1,2,3\}$. For $\hat\ell=4$ with $\mathcal K_{{\hat\ell}}=\{2,5\}$, Step 1 yields $4, 12, 20, 28$ possible delays corresponding to indices $\beta=1,2,3,4$, respectively. Steps 2 and 3 result in $(4,5), (12,2), (20,2), (28,5)$ possible delay-Doppler pairs. The coarse estimate in the first step introduces spurious path $(28,5)$, which needs to be corrected in later steps.
Step 4 takes the $|{\mathcal K}_{\hat \ell}|=2$ pairs with largest correlation, yielding $\mathcal I=\{1,3\}$, resulting in the first and third pairs, i.e., $(4,5), (20,2)$, as final estimates of paths 4 and 6, respectively, in Step 5.  The channel gains of these paths are computed. Steps 1-5 are repeated for each $\hat\ell\in \mathcal J$. $\rm MSE$ is computed in Step 7, where ${\rm MSE}\geq\gamma\sigma_w^2$ due to the inaccurate estimation.
\noindent\textbf{Refinement Steps:}
Refinement steps (see \textit{Refine1} $\&$ \textit{Refine2} functions in Appendix) enable to correct some residual errors from chirp correlation CE (\textit{ChirpCorr} function), which are rarely required as shown in  Table~\ref{alg_freq}. 
\subsubsection{Error correction in Doppler when $|\mathcal K_{{\hat\ell}}|>1$}
For multiple paths having
the same aliased delay with different Doppler shifts, if the following condition
\begin{equation}
    \label{EQ: Refine_1}
    \frac{|{\mathbf C}[\beta^*,\lambda^*_{\beta^*}] - {\mathbf C}[\beta^*,\lambda]|}{|{\mathbf C}[\beta^*,\lambda^*_{\beta^*}]| }\le \epsilon_1 ~~~{\text{for}}~~\lambda^*_{\beta^*}\neq \lambda~~\beta^*\in\mathcal I
\end{equation}
is met, where $\epsilon_1>0$ is a small real number,
then the choice of Doppler shift $\hat k_{\lambda^*_{\beta^*}}({\hat\ell})$ may be incorrect. 
Hence, 
 for such delays $\hat {\textbf l}_{p}=\{\hat l(\hat {\ell},\hat b_{\beta^*})| \beta^*\in \mathcal I\}$, we form all possible delay-Doppler vector pairs $(\hat {\textbf l}_{p}, \hat {\textbf k}_{\text{tmp}})$ for $\hat {\textbf k}_{\text{tmp}}\in \boldsymbol \pi(\mathcal K_{\hat\ell})$, where $\boldsymbol \pi(\mathcal K_{\hat\ell})$ is the list containing all the  permutations of the elements of set $\mathcal K_{\hat\ell}$. Then, for each vector pair $(\hat {\textbf l}_{p}, \hat {\textbf k}_{\text{tmp}})$, we compute the associated channel coefficients $\hat {\textbf h}_{\text{tmp}}$ using (\ref{hi_est_td}) and select the pairing resulting in the least MSE (see also \textit{Refine1} function in Appendix). The path list $\mathcal H$ is updated with the Doppler shift and channel gain of the newly selected pair.
\subsubsection{Error correction when path is unrecognized due to same Doppler} When multiple paths have the same aliased delay with the same Doppler, we estimate only a single path by using the first-stage CE and \textit{ChirpCorr} function, resulting in ${\rm{MSE}}\geq \gamma\sigma_w^2$. This error can be resolved by selecting $\beta\in [1,|{\mathcal B}|]$, satisfying
\begin{equation}\label{EQ: Refine_2}
\frac{|{\mathbf C}(\beta^*,\lambda^*_{\beta^*}) - {\mathbf C}(\beta,\lambda^*_\beta)|}{|{\mathbf C}[\beta^*,\lambda^*_{\beta^*}] |}\le \epsilon_1 ~~~{\text{for}}~~\beta^* \neq \beta.~~
\end{equation}
 Then we pair the associated delay $\hat l = \hat l(\hat{\ell},\hat b_\beta)$ with the Doppler shift $\hat k = \hat k_{\lambda^*_{\beta}}({\hat\ell}) \in \mathcal K_{{\hat\ell}}$ as the delay-Doppler estimates of this path. Note that $\hat k_{\lambda^*_{\beta^*}}(\hat \ell)=\hat k_{\lambda^*_{\beta}}(\hat \ell)$ due to the same Doppler. 
We compute the channel gains using (\ref{hi_est_td}). Moreover, the previously estimated channel gain of  path with delay-Doppler $\{\hat l(\hat{\ell},\hat b_{\beta^*}),\hat k_{\lambda^*_{\beta^*}}(\hat \ell)\}$ is also recomputed using (\ref{hi_est_td}). 
If the refined estimate reduces ${\rm{MSE}}$ then we consider the estimate to be accurate and is appended to path list $\mathcal H$ (see also \textit{Refine2} function in Appendix)

\textit{Example 4:}
Let us continue with Example 3 and assume that the Doppler estimate of paths 4 and 6 in Fig. \ref{over_ch_dd} are incorrect. Then we need to use the first refinement step to correct the Doppler. If the normalized correlation difference for path 4 at both Doppler values  $\{2,5\}$ satisfies (\ref{EQ: Refine_1}), then we correct the pair $(4,5)$ to $(4,2)$. Similarly, we correct the pair $(20,2)$ to $(20,5)$ of path 6, leading to reduced MSE.

\textit{Example 5:}
Continuing with Example 3 and assuming path 5 has not been identified yet, we need to use the second refinement step to choose pairs with stronger correlation satisfying (\ref{EQ: Refine_2}), from the non-selected delay-Doppler pairs $(12,2)$ and $(28,5)$. Hence, we select $(12,2)$ and discard the spurious path $(28,5)$ which has a lower correlation.
     \subsection{Complexity}
     The complexity of the two-stage CE is evaluated in terms of the number of complex multiplications. The primary complexity of the first stage CE involves computing threshold $\mathcal T_{\ell}$ in (\ref{EQ:avg-Power}) which requires $MN$ complex multiplications. In the second stage CE, the complexity is mainly from \textit{ChirpCorr} function including i) conversion from DD to time-domain with complexity $MN\text{log}_2N$, ii) computation of cross-correlation in (\ref{EQ:Cross_correlation}) and (\ref{EQ:cross-corr}) with complexity $MN(l_{\max}+1)+MN|\mathcal B||\mathcal K_{\hat\ell}||{\mathcal J}|$, iii) reconstruction of $\hat {\textbf r}_{\rm t}$ and obtaining $\rm{MSE}$  with a complexity $3MNL+MN$. Hence, the overall complexity order involved in first and second stage is 
     $\mathcal O\Big(MN\Big(\text{log}_2N+|\mathcal B||\mathcal K_{\hat\ell}||{\mathcal J}|+l_{\max}\Big)\Big)$. Further, the major complexity  in the refinement stage lies in the reconstruction of $\hat{\textbf r}_{\rm t}$ and computing $\rm{MSE}$  involving a complexity of $3MNL+MN$, which happen rarely (see Table II). 
     
     \section{MRC Detection for OTFS in Overspread Channels}\label{Sec:Detection}
     Zero padded OTFS (ZP-OTFS) is spectrally inefficient for an overspread channel with long delays. Hence, we consider reduced zero padded OTFS (RZP-OTFS) \cite{tharaj_universal_MRC}. We recall the DT domain input-output relation of RZP-OTFS for the overspread channel  from \cite{tharaj_universal_MRC} as
     \begin{equation} \label{DT_Relation_overspread}
         \tilde{\textbf y}_m=\sum_{i=0}^{L-1} \tilde{\boldsymbol \nu}_{m,l_i}\circ \Big(\boldsymbol{\Pi}_{l,N}^{-\lfloor \frac{m-l_i}{M}\rfloor} \tilde{\textbf x}_{[m-l_i]_M}\Big)+ \tilde{\textbf w}_m, 
     \end{equation}
     where $\tilde{\textbf y}_m \in \mathbb{C}^{N\times 1}$,  $m\in[0, M-1]$,  are the DT domain received OTFS symbol vectors at $m$-th delay. The  $\tilde{\textbf w}_m \in \mathbb{C}^{N\times 1}$ is the AWGN vector, and $\tilde{\boldsymbol \nu}_{m,l_i}\in\mathbb{C}^{N\times 1}$ is the DT domain channel vector with elements given as
     \begin{equation}
         \tilde{\boldsymbol \nu}_{m,l_i}[n]=
   h_ie^{\jmath\frac{2\pi}{MN}k_i(m-{l_i})}
   e^{\jmath\frac{2\pi}{N}k_i n},
     \end{equation} 
   for $n\in[0, N-1]$.      Further, 
     \begin{equation}
         \boldsymbol{\Pi}_{l,N}=\begin{bmatrix}0 & \cdots&0&0 \\1 & \cdots&0&0\\\vdots&\ddots&\ddots&\vdots\\0&\cdots&1&0 \end{bmatrix}
     \end{equation}
     is a {\em linear shift} matrix. 
     Following the same notations in \cite{tharaj_universal_MRC} (subscript $i$ in channel parameters are omitted in the following), we adopt the low complexity DT domain MRC detector for RZP-OTFS with overspread delay. The MRC detection involves extracting and coherently combining the desired signal components $\tilde {\textbf x}_m$ received from all diversity branches from the multipath effect to  improve the
signal to interference plus noise ratio (SINR) in each iteration. To observe the desired signal and interference components in each branch, the input-output relation in (\ref{DT_Relation_overspread}) can be rewritten as
\begin{align}\label{DT_desi_intf_eqn}
\tilde{\textbf y}_{m+l}= \tilde{\boldsymbol \nu}_{[m+l]_M,l}\circ \Big(\breve{\boldsymbol \Pi}^{-\lfloor\frac{M-m-l}{M}\rfloor} \tilde{\textbf x}_{m}\Big)+ \sum_{l^\prime\in\mathcal L, l^\prime\neq l} \tilde{\boldsymbol \nu}_{[m+l]_M,l^{\prime}}\circ \Big(\breve{\boldsymbol \Pi}^{-\lfloor\frac{M-m-l}{M}\rfloor} \tilde{\textbf x}_{[m+l-l^\prime]_M}\Big)+ \tilde{\textbf w}_m. 
\end{align}
Based on (\ref{DT_desi_intf_eqn}), we formulate the DT MRC detector following \cite{tharaj_rake_MRC_journal, tharaj_universal_MRC} and is shown in Algorithm \ref{Alg4}. Here, we let $\breve{\boldsymbol{\Pi}}=\boldsymbol{\Pi}^T_{l,N}$, $\mathcal D (\cdot)$ represents the hard decision QAM symbol demodulator, $\hat{\tilde{\textbf x}}_m^{(0)}$ denote the initial estimate of $\tilde{\textbf x}_m$,  the output $\hat{\bf x}_m$ denotes the final estimate of ${\bf x}_m$, and $N_{\text{iter}}$ is the total number of iterations. For details of Algorithm \ref{Alg4}, we refer readers to \cite{tharaj_universal_MRC}. Due to the use of RZP-OTFS in an overspread channel, the modified MRC detection differs from  \cite{tharaj_universal_MRC} in the linear shift matrix and DT channel vector. In particular, the linear shift matrix is modified to  $\breve{\boldsymbol \Pi}^{-\lfloor\frac{M-m-l}{M}\rfloor}$ from $\boldsymbol{\Pi}_{l,N}^{\lfloor \frac{m-l}{M}\rfloor}$ and DT channel vector $\tilde{\boldsymbol \nu}_{m+l,l}$ is modified to $\tilde{\boldsymbol \nu}_{[m+l]_M,l}$.  This modification is in accordance with (\ref{DT_desi_intf_eqn}) to capture the desired DT symbols in the overspread paths. 
          \begin{algorithm}[t]
    		\caption{Iterative DT MRC Detector for RZP-OTFS}
    		\begin{algorithmic} [1]
    			\STATE {\textbf{Input:} $\hat{\tilde{\textbf x}}_m^{(0)}, \tilde{\textbf y}_m, \tilde{\boldsymbol \nu}_{m,l} \hspace{.2cm} \forall l,m$}.
       \STATE{\textbf{Output:} $\hat{{\textbf x}}_m\hspace{.2cm} \forall m$}
       \FOR{$m \leftarrow 0$ to $M-1$}
       \STATE{$\Delta \tilde{\textbf y}^{(0)}_{m}\leftarrow \tilde{\textbf y}_{m}-\sum_{l\in\mathcal L}\tilde{\boldsymbol \nu}_{m, l} \circ\Big(\boldsymbol{\Pi}_{l,N}^{-\lfloor \frac{m-l}{M}\rfloor} \hat{\tilde{\textbf x}}^{(0)}_{[m-l]_M}\Big)$}
       \STATE{$ \tilde{\textbf d}_m\leftarrow\sum_{l\in \mathcal L}\breve{\boldsymbol \Pi}^{-\lfloor\frac{M-m-l}{M}\rfloor}\Big(\tilde{\boldsymbol \nu}^*_{[m+l]_M,l} \circ \tilde{\boldsymbol \nu}_{[m+l]_M,l}\Big)$}
    		\ENDFOR
    		\FOR{$i \leftarrow 1$ to $N_{\text{iter}}$}
      \FOR{$m \leftarrow 0$ to $M-1$}
      \STATE{$\Delta \tilde{\textbf y}^{(i)}_{m}\leftarrow\Delta \tilde{\textbf y}^{(i-1)}_{m}$}
      \ENDFOR
    		\FOR{$m \leftarrow 0$ to $M-1$}
      \STATE{$\Delta \tilde{\textbf g}_m^{(i)}\leftarrow  \sum_{l\in \mathcal L}\Big(\breve{\boldsymbol \Pi}^{-\lfloor\frac{M-m-l}{M}\rfloor}\Big(\tilde{\boldsymbol \nu}^*_{[m+l]_M,l}\Big) \circ\breve{\boldsymbol \Pi}^{-\lfloor\frac{M-m-l}{M}\rfloor}\Delta \tilde{\textbf y}^{(i)}_{[m+l]_M}\Big)$}
      \STATE{$  \tilde{\textbf c}_m^{(i)}\leftarrow\hat{\tilde{\textbf x}}^{(i-1)}_{m}+\Delta \tilde{\textbf g}_m^{(i)}\oslash\tilde{\textbf d}_m,$}
      \STATE{$\hat{\tilde{\textbf x}}^{(i)}_{m}\leftarrow\bar\delta\textbf F_N^H \cdot \mathcal D(\textbf F_N \tilde{\textbf c}_m^{(i)})+(1-\bar\delta)\tilde{\textbf c}_m^{(i)},\hspace{.2cm} 0\leq \bar\delta\leq 1$}
      \FOR{$l \in \mathcal L$}
    		\STATE{$ \Delta \tilde{\textbf y}^{(i)}_{[m+l]_M}\leftarrow\Delta \tilde{\textbf y}^{(i-1)}_{[m+l]_M}-\tilde{\boldsymbol \nu}_{[m+l]_M, l} \circ\Big(\breve{\boldsymbol \Pi}^{-\lfloor\frac{M-m-l}{M}\rfloor}(\hat{\tilde{\textbf x}}^{(i)}_{m}-\hat{\tilde{\textbf x}}^{(i-1)}_{m})\Big)$}
      \ENDFOR
    		\ENDFOR
      \IF { $ ||\Delta\tilde{\textbf y}_m^{(i)}||\geq||\Delta\tilde{\textbf y}_m^{(i-1)}||, \hspace{.2cm}\forall m$} 
      \STATE{\textbf{EXIT}}
      \ENDIF
    		\ENDFOR
    		\end{algorithmic} \label{Alg4}
    	\end{algorithm}
\subsection{Detection Complexity and Convergence}
The initialization step in Algorithm \ref{Alg4} in lines 4-5, executed once, requires $2MNL$ complex multiplications. Further, at each iteration, the MRC combining in line 12, DT symbol-vector estimates update in line 13, transformation of  the estimated DT samples to DD and vice versa in line 14, and  update of $\Delta \tilde{\textbf y}_{[m+l]_M}$ in line 16 requires $MN((2L+1)+2\text{log}_2N)$ complex multiplications \cite{tharaj_universal_MRC}. Hence, the overall complexity order is $\mathcal O(MNL)$, which is less than the standard MPA complexity given by $\mathcal O(MNLQ)$. 

The convergence of the MRC detector for RZP-OTFS can be extended straightforwardly from that of the MRC detector for ZP-OTFS in \cite{tharaj_rake_MRC_journal}.


\section{Results and Discussions}\label{Sec:Simulation}
 \begin{table}[t]
 \caption{\small Simulation Parameters}
\begin{center}
\begin{tabular}{ | m{5em} | m{4.5em} | m{4.5em}|m{4.7em}|} 
  \hline
   \vspace{0.1 cm}Parameter & \vspace{0.1 cm}Channel A  &\vspace{0.1 cm} Channel B &\vspace{0.1 cm} Channel C\\[1ex]\hline\hline
  \vspace{0.1 cm} $\Delta f$&\vspace{0.1 cm} 15 KHz&\vspace{0.1 cm} 15 KHz&\vspace{0.1 cm} 900 KHz\\[1ex]\hline
  \vspace{0.1 cm}$M\times N$& \vspace{0.1 cm}$512\times 128$&\vspace{0.1 cm}$512\times 128$&\vspace{0.1 cm}$512\times 128$ \\ [1 ex]
  \hline
  \vspace{0.1 cm}Speed& \vspace{0.1 cm}500 kmph&\vspace{0.1 cm} 500 kmph&\vspace{0.1 cm}1000 kmph\\[1ex]\hline
  \vspace{0.1 cm}$L$ (Paths)&\vspace{0.1 cm}$9$&\vspace{0.1 cm} $9$&\vspace{0.1 cm} $9$\\[1ex]\hline
  \vspace{0.1 cm} $|h_i|^2$ profile &Uniform&EVA&ETU\\[1ex]\hline
  \vspace{0.1 cm}$l_i$ profile&$\mathcal U[0,l_{\max}]$&$\mathcal U[0,l_{\max}]$&ETU\\[1ex]\hline
  \vspace{0.1 cm}$\l_{\max}$&\vspace{0.1 cm}2400&\vspace{0.1 cm}2400&\vspace{0.1 cm}2400\\[1ex]\hline
  \vspace{0.1 cm}$k_{\max}$&\vspace{0.1 cm}16&\vspace{0.1 cm}16&\vspace{0.1 cm}1\\[1ex]\hline
\end{tabular}\label{table_chann_param}
\end{center}
\end{table}
\begin{figure}
  \centering
	\includegraphics[width=.5\linewidth]{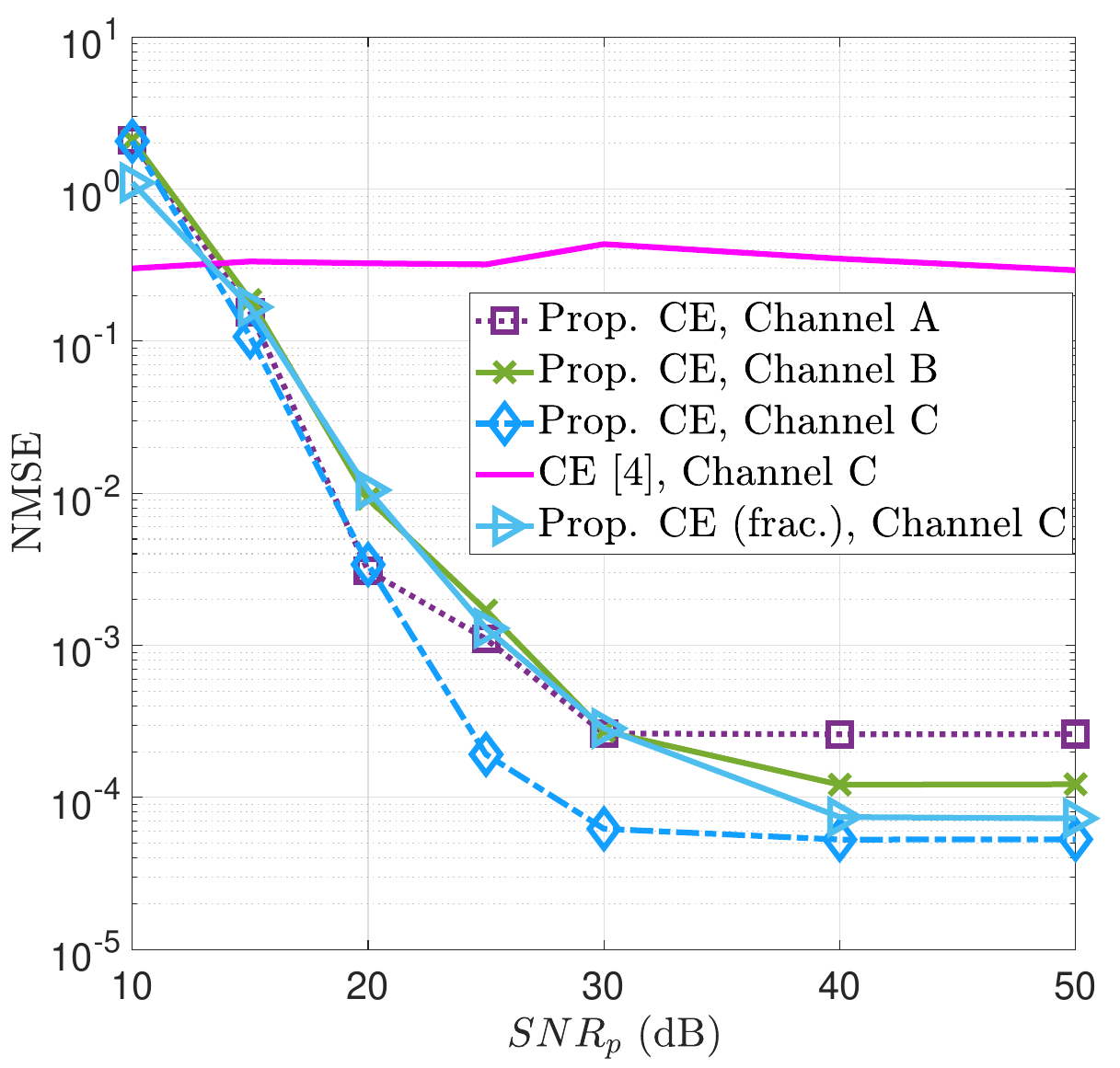}%
	\caption{\small NMSE vs $\text{SNR}_{\rm p}$.}
	\label{fig:fig11}
\end{figure}%
 \begin{figure}
\centering
\begin{minipage}{.5\textwidth}
  \centering
\includegraphics[width=.9\linewidth]{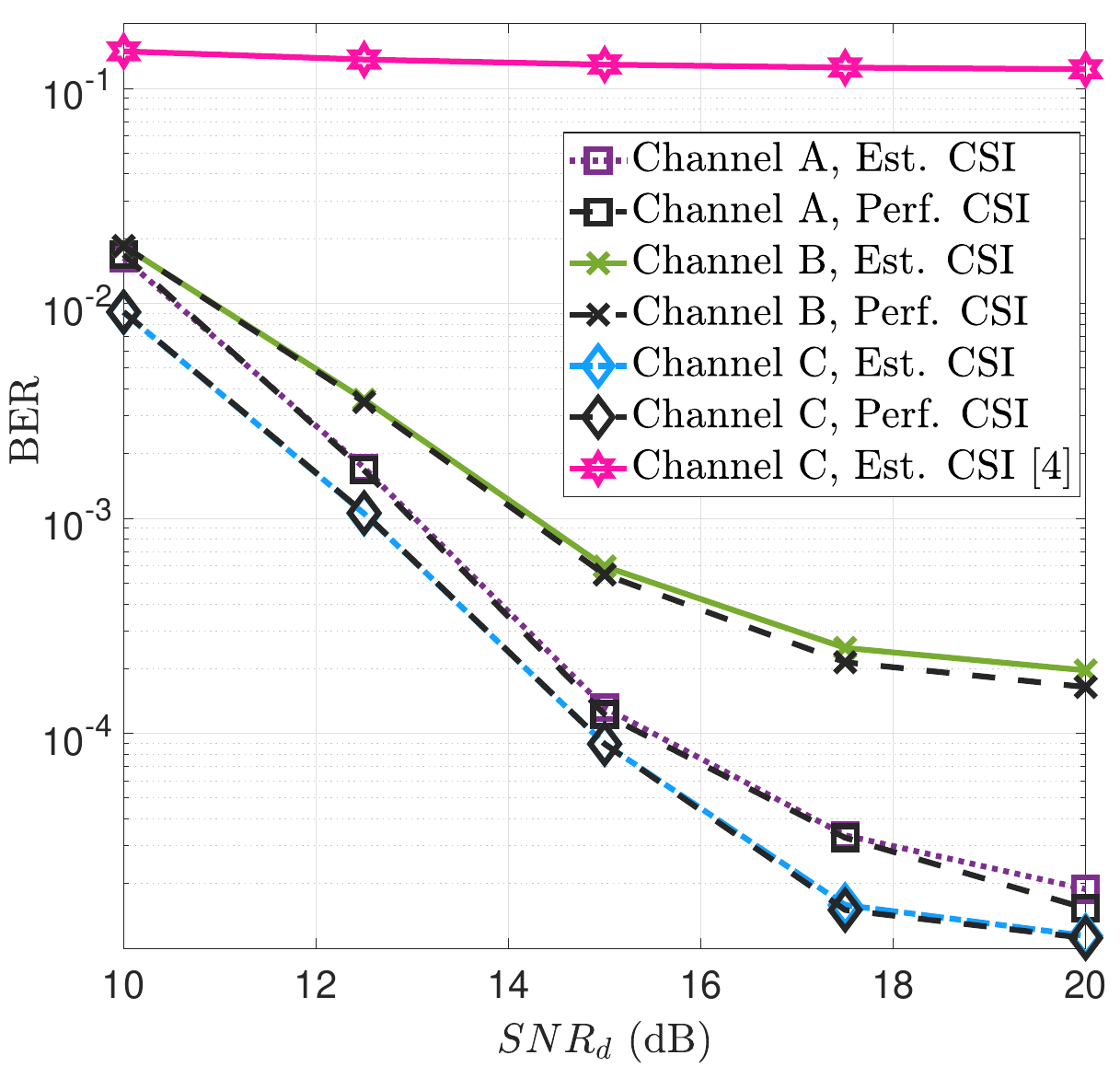}
	\caption{\small  BER vs $\text{SNR}_{\rm d}$, for $\text{SNR}_{\rm p}=30$\,dB and $\text{SNR}_{\rm c}=23$\,dB using the proposed CE and MRC detection.}
	\label{fig:fig12}
\end{minipage}%
~~
\begin{minipage}{.5\textwidth}
  \centering
\includegraphics[width=.96\linewidth]{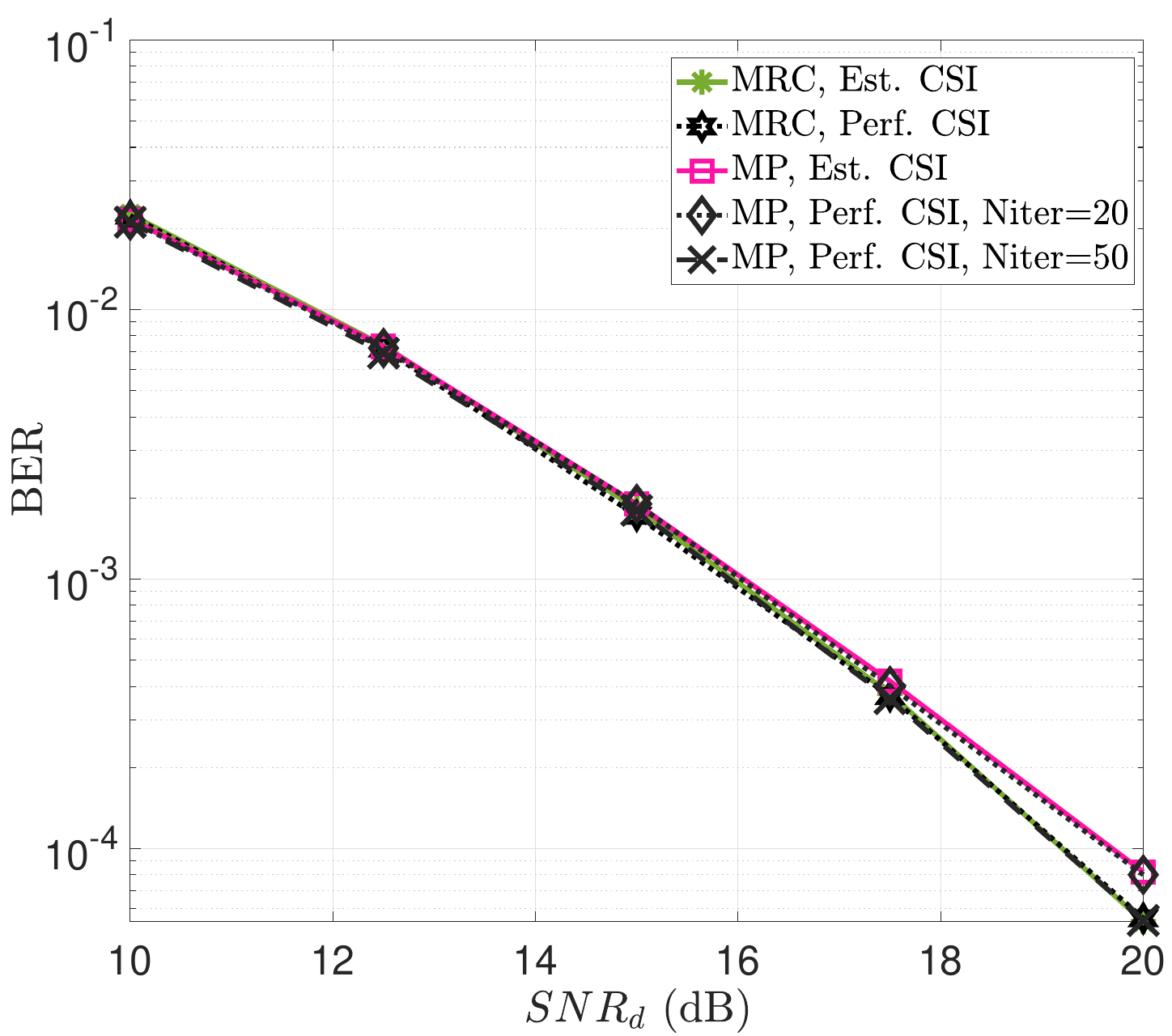}%
	\caption{\small BER vs $\text{SNR}_{\rm d}$ comparison between MRC and MP $(M=N=32, L=4$, 4-QAM).}
	\label{ber_comp_MRC_MP}
\end{minipage}
\end{figure}
In this section, we illustrate the performance of the proposed two-stage CE in terms of NMSE and BER, where the NMSE is defined as \[\text{NMSE}=\frac{\sum_{m=0}^{M-1}\sum_{i=0}^{L-1}||\widehat{\widetilde{\boldsymbol \nu}}_{m,l_i}-\tilde{\boldsymbol \nu}_{m,l_i}||^2}{\sum_{m=0}^{M-1}\sum_{i=0}^{L-1}||\tilde{\boldsymbol \nu}_{m,l_i}||^2},\]
where $\widehat{\widetilde{\boldsymbol \nu}}_{m,l_i}$ denotes the estimated DT domain channel vector. Here we denote the pilot, chirp, and data SNRs in the time domain by $\text{SNR}_{\rm p}=\frac{|x_{\rm p}|^2}{N\sigma_w^2}$, $\text{SNR}_{\rm c}=\frac{2A^2}{\sigma_w^2}$, and $\text{SNR}_{\rm d}=\frac{E_s}{\sigma_w^2}$, respectively. The performance of the overspread MRC detector is evaluated utilizing the estimated channel parameters from the training frame with the assumption that it remains unchanged between the training and data frames.

All the relevant simulation parameters are summarized in Table~\ref{table_chann_param}. We first adopt two synthetic overspread channel models: Channels A and B,  both having at least initial two underspread paths. Further, Channel A's paths follow uniform power distribution while Channel B's paths follow EVA model power profile \cite{ETU}. We then adopt a practical channel model, Channel C, which uses the ETU model \cite{ETU} with an overspread delay. The Doppler shifts are generated using the Jakes formula for Channels A, B, and C. In our simulation, we set the different parameters as $\alpha=4, \delta=30, \alpha^\prime=2, \Gamma=\text{SNR}_{\rm p}$, $ \Gamma^\prime=500$, $\gamma=2$, and $\epsilon_1=0.6$. These parameters can be chosen according to the channel models and system parameters to optimize the performance. 


Fig.~\ref{fig:fig11} illustrates the NMSE performance as a function of pilot $\text{SNR}_{\rm p}$ for chirp $\text{SNR}_{\rm c}=23$\,dB. For all the channel types, we observe the NMSEs decrease significantly as pilot SNR increases up to $30$\,dB and floor beyond $30$ dB. This indicates, for such settings, pilot SNR should be at least $7$ dB higher than chirp SNR for reliable performance.   Furthermore, we observe that the embedded pilot CE \cite{embedded_pilot_raviteja} designed for underspread channels, completely fails to accurately estimate the overspread channels.

We also evaluate the performance of the proposed CE for the fractional delay and integer Doppler case for Channel C in the figure. For fractional delay $l_i$, the discrete delay-time channel response for a channel tap $p$ at time $qT_s$ is \cite{PP_Oversam_MRC}
 \begin{equation}
   h[q,p]=h(qT_s,p) =\sum_{i=0}^{L-1}h_ie^{j\frac{2\pi}{MN}k_i(q-l_i)}\text{Sinc}(p-l_i),
 \end{equation}
 where $T_s$ is sampling interval. Here, we only consider the channel taps $p \in {\cal P}$, where
\begin{equation}
    \mathcal P=\left\{p: |\text{Sinc}(p-l_i)|>\epsilon\right\}
\end{equation}
is the set containing the dominant samples of $\text{Sinc}(p-l_i)$ greater than a threshold $\epsilon$.

We employ our proposed estimator to estimate the channel taps and Doppler shifts. It can be observed that the proposed estimator provides reliable performance in the presence of fractional delay.

In Fig. \ref{fig:fig12}, we illustrate the BER performance of the RZP-OTFS system with MRC detector that utilizes the proposed channel estimator. We set the pilot SNR to 30\,dB and chirp SNR to 23\,dB. Further, information symbols are modulated using 4-QAM and $N_{\text{iter}}$ for detection is set to 5. The weighting coefficient $\bar\delta$ is set to 1. The MRC detector integrated with the proposed channel estimator can attain a BER of $2\times 10^{-4}$ at an SNR of about 14.5\,dB, 20\,dB, and 14.1\,dB for channels A, B, and C, respectively.  In contrast, we can observe that the performance degrades significantly while utilizing the aliased delay and Doppler estimates \cite{embedded_pilot_raviteja} for channel C. Further, we can observe that the performance with the estimated channel degrades marginally compared to the perfectly known channel at all SNRs which solidifies the reliability of the proposed overspread channel estimator. 

Table \ref{alg_freq} shows the percentage of times the refinement steps are invoked. We observe that  \textit{Refine1} function, rectifying any incorrect Doppler estimates, is invoked $2 \%$ of the time for both channels A and B, sharing same delay and Doppler distribution. 

Further, the invocation of \textit{Refine2} function depends on occurrence frequency of delay ambiguity. For channels A and B, the algorithm is invoked at same rate of $0.25\%$, due to their identical delay and Doppler distributions. 
Channel C does not require any refinement algorithm since all the paths have distinct aliased delays. We conclude that the proposed {\em two-stage channel estimation without refinement steps}  can estimate the channel parameters for most channel realizations and in some rare cases only, refinement steps are required. Thus, the additional complexity in refinement steps is minimal.

In Fig. \ref{ber_comp_MRC_MP}, we compare the BER performance of MRC with MP detection under an overspread channel setting. Due to the larger memory requirement in MP, we assessed the performance using lower frame dimensions ($M=N=32$) under a synthetic channel configuration comprising four paths, with two characterized as underspread and the remaining as overspread. The channel gains exhibit a uniform distribution. The maximum iteration for the MP and MRC is 20. Further, the damping factor for MP is set to 0.125 and weighting coefficient $\bar\delta$ for MRC is 0.25. We employed the proposed CE for both the detectors. We can observe comparable performance between the MP and MRC detectors at SNR below 18 dB. At higher SNR levels (beyond 18 dB), MRC exhibits 0.3 dB SNR gain to attain BER of $10^{-4}$.


To achieve a comparable performance to the MRC detection with 20 iterations,  we need to increase the number of iterations of MP from 20 to 50 with an optimized damping factor of 0.06, as shown Fig. \ref{ber_comp_MRC_MP}.  
In summary, it is shown that MRC detection can offer at least comparable error performance to MP at a much lower complexity.

Additionally, both detectors exhibits similar performance when utilizing estimated channel information compared to the ideal scenario of perfect channel information.
 \begin{table}[t]
 \caption{\small  Refinement Algorithm Invocation Percentage}
\begin{center}
\renewcommand{\arraystretch}{1.6}
\begin{tabular}{ | m{6.3em} | m{4.29em} | m{4.25em}| m{4.25em}|} 
  \hline
   \multirow{2}{*}{~~~~~~Type } & \multicolumn{3}{c|}{~Percentage} \\[1ex]\cline{2-4}
    &Channel A&Channel B&Channel C\\[1ex]\hline\hline
  \vspace{0.1 cm} ~~~Func. \textit{Refine1}&\vspace{0.1 cm}  ~~~~~$2\%$&~~~~~$2\%$&~~~~~$0\%$\\[1ex]\hline
  \vspace{0.1 cm} ~~~Func. \textit{Refine2}&\vspace{0.1 cm} ~~~~$0.25\%$&~~~~$0.25\%$&~~~~~$0\%$ \\[1ex]\hline
\end{tabular}\label{alg_freq}
\end{center}
\end{table}




\section{Conclusion}\label{Sec:Concl}
In this paper, we study the CE scheme and detection method for an OTFS system in  the overspread channel exhibiting a very long delay spread exceeding the block duration in a frame. We first propose a two-stage CE method using a DD domain training frame. This frame consists of a dual chirp converted from time-domain and a higher power pilot. In the first stage, we employ a DD domain embedded pilot CE using adaptive thresholds to estimate aliased delays and Doppler shifts, followed by identifying all the underspread paths not coinciding with any overspread ones. In the second stage, by leveraging the time-domain correlation between the transmit and received dual chirps, we estimate actual delays and Doppler shifts of the remaining paths. In rare cases, refinement steps can be used to resolve delay ambiguity and rectify incorrect Doppler estimation among the paths sharing the same aliased delays. Furthermore, we present a modified low-complexity MRC detection algorithm for OTFS in overspread channels. We also discuss the complexity of the proposed CE and the modified MRC detection, as well as the detection convergence. 
Finally, we provide simulation results, which demonstrate the effectiveness of the proposed CE scheme and the MRC detection algorithm for overspread channels. Extension of our proposed scheme in presence of fractional Doppler will be investigated in our future work.
\newpage
\appendix  \label{functions}
\vspace{4mm}
\hrule
\begin{center}
\vspace{0.2 cm}
 {\bf{ Stage 2 CE Chirp Correlation}} 
 \end{center}	
 \hrule
 \begin{algorithmic} [1]
    \STATE{$\text{\bf func}~[{\mathcal H}]=\textit{ChirpCorr} (\textbf r_{\rm t}, \textbf s_{\rm t}, \gamma, \sigma_w^2,\Gamma,\Gamma^\prime, l_{\max},i,{\mathcal J}, \mathcal H,$$ \mathcal K_{\hat \ell}, \forall  \hat\ell\in \mathcal J)$}
     \STATE {Set $\mathcal B=\{\}$}
      \FOR{$q\rightarrow0~\text{to}~MN-1$}

      \STATE{$\check{\textbf r}_{\rm c}[q]\leftarrow \textbf r_{\rm t}[q]$}
      
      \IF{$|\textbf r_{\rm t}[q]|^2>\Gamma$}
\STATE{$\check{\textbf r}_{\rm c}[q]\leftarrow 0$}
       \ENDIF
       \ENDFOR
       \STATE{Obtain $\textbf p[0,0,q], q\in[0,MN-1]$ using (\ref{EQ:generic_Chirp})}
       \FOR{$q\rightarrow0~\text{to}~l_{\max}$}
       \STATE{Compute cross-correlation $R_{\check {\textbf r}_{\rm c},\textbf p}[q]$ using (\ref{EQ:Cross_correlation})}
       \IF {$|R_{\check {\textbf r}_{\rm c},\textbf p}[q]|\geq\Gamma^\prime$}
       \STATE{Obtain $\hat b_\beta=\lfloor{\frac{ q}{M}}\rfloor$}
       \STATE{$\mathcal B\leftarrow\mathcal B\cup\hat b_\beta$}
       \ENDIF
       \ENDFOR
        \FOR{$\hat \ell\in {\mathcal J}$}
    		\FOR{${\beta}\rightarrow1$ to $|\mathcal B|$ } 
      \STATE {Obtain
        $\hat l(\hat {\ell},\hat b_\beta)=\hat {\ell}+\hat b_\beta M, ~~\hat b_\beta\in \mathcal B$}
      \FOR{$\lambda\rightarrow1$ to $|\mathcal K_{{\hat\ell}}|$}
    		\STATE{ Obtain ${\textbf p}[0, \hat k_{\lambda}({\hat\ell}), q]$ using (\ref{EQ:generic_Chirp})}
      \STATE{ Compute cross-correlation $\textbf{C}[\beta,\lambda]$ using (\ref{EQ:cross-corr})}
      \ENDFOR
      \STATE{Find $(\beta,\lambda^*_\beta)=\arg\max\limits_{\lambda} \textbf C[\beta,\lambda]$}
      \STATE{Obtain delay-Doppler pair $(\hat l(\hat {\ell},\hat b_\beta), \hat k_{\lambda_{\beta}^*}({\hat\ell}))$}
    \ENDFOR
    \STATE{Take $|{\mathcal K}_{\hat \ell}|$ largest values in $\textbf C[\beta,\lambda^*_\beta]$ with index set $\mathcal I$}
    \STATE{Determine $\mathcal H^\prime\leftarrow\Big\{\Big(\hat l_i=\hat l(\hat {\ell},\hat b_{\beta^*}),\hat k_i=\hat k_{\lambda^*_{\beta^*}}({\hat\ell}), \hat h_i\Big)\Big|\beta^*\in\mathcal I\Big\}$}
     \STATE{$\mathcal H\leftarrow\mathcal H \cup \mathcal H^\prime$.}
     \STATE{$i\leftarrow i+|\mathcal K_{\hat \ell}|$}
      \ENDFOR
      \STATE{ ${\rm MSE}\leftarrow\mathcal E\{\textbf r_{\rm t}, \hat{\textbf r}_{\rm t}\}$.}
      \IF{${\rm MSE}\geq\gamma\sigma_w^2$}
       \FOR{$\hat \ell\in {\mathcal J}$}
       \IF{$|\mathcal K_{{\hat\ell}}|>1$}
       \vspace{0.05 cm}
      \STATE{$[{\rm MSE, \mathcal H}]=\textit{Refine}1 {(\rm{MSE}}, \textbf r_{\rm t}, \textbf s_{\rm t}, \mathcal K_{{\hat\ell}}, \mathcal I, {\bf C}, \mathcal H, \epsilon_1)$}
      \ENDIF
      \ENDFOR
      \ENDIF
       \IF{${\rm MSE}\geq\gamma\sigma_w^2$}
       \FOR{$\hat \ell\in {\mathcal J}$}
       \STATE{$[{\rm MSE, \mathcal H},i]=$$\textit{Refine}2 {(\rm{MSE}}, \textbf r_{\rm t}, \textbf s_{\rm t}, \mathcal K_{{\hat\ell}},\mathcal B, \mathcal I, {\bf C}, \mathcal H, \epsilon_1,i)$}
      \ENDFOR
      \ENDIF
      \STATE{Return $\mathcal H$}
    		\end{algorithmic} \label{Alg_chirp}
       \hrule
\vspace{1.2 cm}
\hrule
\begin{center}
    		\bf{Refinement Step 1 (to rectify Doppler)}
      \end{center}
      \hrule
    		\begin{algorithmic} [1]
  \STATE{$\text{\bf func}~[{\rm MSE, \mathcal H}]=\textit{Refine1} {(\rm{MSE}}, \textbf r_{\rm t}, \textbf s_{\rm t}, \mathcal K_{{\hat\ell}}, \mathcal I, {\bf C}, \mathcal H, \epsilon_1)$}
      \IF{ $\frac{|{\mathbf C}[\beta^*,\lambda^*_{\beta^*}] - {\mathbf C}[\beta^*,\lambda]|}{|{\mathbf C}[\beta^*,\lambda^*_{\beta^*}]|}\le \epsilon_1 ~{\text{for any}}~\beta^* \in {\mathcal I}, {\text{and}}~\lambda\neq \lambda^*_{\beta^*} $}
   \STATE{Find delay $\hat{\textbf{l}}_{p}=\{l(\hat\ell,\hat b_{\beta^*})|\beta^*\in\mathcal I\}$ corresponding to $\hat\ell$ in $\mathcal H$}
  \STATE{Generate all permutations  of $\mathcal K_{\hat\ell} $ to form the list $\boldsymbol{\pi}(\mathcal K_{\hat\ell})$}
\FOR{$\hat{\textbf{k}}_{\text{tmp}}\in \boldsymbol{\pi}(\mathcal K_{\hat\ell})$}
\STATE{Associate $\hat{\textbf{k}}_{\text{tmp}}$ with $\hat{\textbf{l}}_{{p}}$ and compute $\hat {\textbf h}_{\text{tmp}}$ using (\ref{hi_est}). }
   \STATE{Compute $\mathcal E\{\textbf r_{\rm t}, \hat{\textbf r}_{\rm t}\}$ with updated $(\hat{\textbf{l}}_{{p}},\hat {\textbf k}_{\text{tmp}},\hat {\textbf h}_{\text{tmp}})$ .}
    \IF{ $\mathcal E\{\textbf r_{\rm t}, \hat{\textbf r}_{\rm t}\} <\rm{MSE}$}
            \STATE{$(\hat {\textbf l}_{p},\hat {\textbf k}_{p},\hat {\textbf h}_{p})\leftarrow(\hat {\textbf l}_{p},\hat {\textbf k}_{\text{tmp}},\hat {\textbf h}_{\text{tmp}})$}
                  \STATE{$\rm{MSE} \leftarrow  \mathcal E\{\textbf r_{\rm t}, \hat{\textbf r}_{\rm t}\}$}
                                   \ENDIF
\ENDFOR
       \ENDIF
\STATE{Return $\mathcal H, {\rm MSE}$}
    		\end{algorithmic} \label{refine1}
      \hrule
    \vspace{1 cm}
\hrule

\begin{center}
    		\bf{Refinement Step 2 (to identify delay)}
      \end{center}
      \hrule
    		\begin{algorithmic} [1]
   \STATE{$\text{\bf func}~[{\rm MSE, \mathcal H},i]=$$\textit{Refine2} {(\rm{MSE}}, \textbf r_{\rm t}, \textbf s_{\rm t}, \mathcal K_{{\hat\ell}}, \mathcal B, \mathcal I, {\bf C}, \mathcal H, \epsilon_1,i)$}
      \FOR{$\beta\rightarrow1~\text{to}~|{\mathcal B}|$}
      \IF{$\frac{|{\mathbf C}(\beta^*,\lambda^*_{\beta^*}) - {\mathbf C}(\beta,\lambda^*_\beta)|}{|{\mathbf C}[\beta^*,\lambda^*_{\beta^*}]|}\le \epsilon_1 ~{\text{and}}~\beta \neq \beta^*$ }
      \STATE{Choose potential delay $\hat l_i=\hat l(\hat{\ell},\hat b_\beta)$ with the Doppler $\hat k_i=\hat k_{\lambda_\beta^*}({\hat\ell})\in\mathcal K_{{\hat\ell}}$.}
      \STATE{${\rm h_{tmp}}\leftarrow$ $\hat h_{{p}}$ computed in (\ref{hi_est}) associated with $\hat l_{p}=\hat l(\hat{\ell},\hat b_{\beta^*}), \hat k_{p}=\hat k_{\lambda^*_{\beta^*}}({\hat\ell})=\hat k_{\lambda^*_\beta}({\hat\ell})$ in $\mathcal H$}
      \IF{$\hat l_{{p}}<\hat l_i$}
      \STATE{Recompute $\hat h_{p}$ in $\mathcal H$ using (\ref{hi_est_td}).}
      \STATE{Compute $\hat h_i$ using (\ref{hi_est_td}).}
      \ELSE
        \STATE{Compute $\hat h_i$ using (\ref{hi_est_td}).}
        \STATE{Recompute $\hat h_{p}$ in $\mathcal H$ using (\ref{hi_est_td}).}
      \ENDIF
      \STATE{Compute $\mathcal E\{\textbf r_{\rm t}, \hat{\textbf r}_{\rm t}\}$ using $\mathcal H\cup\{(\hat l_i,\hat k_i,\hat h_i)\}$}
            \IF{$\mathcal E\{\textbf r_{\rm t}, \hat{\textbf r}_{\rm t}\}<\rm{MSE}$}
             \STATE{$\mathcal H\leftarrow\mathcal H \cup\{(\hat l_i,\hat k_i,\hat h_i)\}$}
     \STATE{$i\leftarrow i+1$}
     \STATE{${\rm MSE}\leftarrow \mathcal E\{\textbf r_{\rm t}, \hat{\textbf r}_{\rm t}\}$}
           \ELSE
      \STATE{$\hat h_{p}\leftarrow {\rm h_{tmp}}$}
      \ENDIF
      \ENDIF
      \ENDFOR
      \STATE{Return $\mathcal H, {\rm MSE}, i$}
    		\end{algorithmic} \label{refine2}
      \hrule

	\ifCLASSOPTIONcaptionsoff
	\newpage
	\fi
	\bibliographystyle{IEEEtran}
	\bibliography{reference}		

\begin{thebibliography}{10}
\providecommand{\url}[1]{#1}
\csname url@samestyle\endcsname
\providecommand{\newblock}{\relax}
\providecommand{\bibinfo}[2]{#2}
\providecommand{\BIBentrySTDinterwordspacing}{\spaceskip=0pt\relax}
\providecommand{\BIBentryALTinterwordstretchfactor}{4}
\providecommand{\BIBentryALTinterwordspacing}{\spaceskip=\fontdimen2\font plus
\BIBentryALTinterwordstretchfactor\fontdimen3\font minus \fontdimen4\font\relax}
\providecommand{\BIBforeignlanguage}[2]{{%
\expandafter\ifx\csname l@#1\endcsname\relax
\typeout{** WARNING: IEEEtran.bst: No hyphenation pattern has been}%
\typeout{** loaded for the language `#1'. Using the pattern for}%
\typeout{** the default language instead.}%
\else
\language=\csname l@#1\endcsname
\fi
#2}}
\providecommand{\BIBdecl}{\relax}
\BIBdecl

\bibitem{hadani2017orthogonal}
R.~Hadani \emph{et~al.}, ``{Orthogonal Time Frequency Space Modulation},'' in \emph{Proc. IEEE Wireless Commun. Net. Conf.}, San Francisco, CA, USA, 2017, pp. 1--6.

\bibitem{delay_Doppler_book}
Y.~Hong, T.~Thaj, and E.~Viterbo, \emph{{Delay-Doppler Communications}}.\hskip 1em plus 0.5em minus 0.4em\relax Elsevier, 2022.

\bibitem{VTC_ChaEst_2018}
P.~Raviteja, K.~T. Phan, Y.~Hong, and E.~Viterbo, ``{Embedded Delay-Doppler Channel Estimation for Orthogonal Time Frequency Space Modulation},'' in \emph{Proc. IEEE 88th Veh. Tech. Conf.}, Chicago, IL, USA, 2018, pp. 1--5.

\bibitem{embedded_pilot_raviteja}
P.~Raviteja, K.~T. Phan, and Y.~Hong, ``{Embedded Pilot-Aided Channel Estimation for OTFS in Delay–Doppler Channels},'' \emph{IEEE Trans. Veh. Tech.}, vol.~68, no.~5, pp. 4906--4917, 2019.

\bibitem{Shen_OMP_2019}
W.~Shen, L.~Dai, J.~An, P.~Fan, and R.~W. Heath, ``{Channel Estimation for Orthogonal Time Frequency Space (OTFS) Massive MIMO},'' \emph{IEEE Trans. Sig. Proc.}, vol.~67, no.~16, pp. 4204--4217, 2019.

\bibitem{Chocks_multi_user_MIMO_ch_est_2020}
O.~K. Rasheed, G.~D. Surabhi, and A.~Chockalingam, ``{Sparse Delay-Doppler Channel Estimation in Rapidly Time-Varying Channels for Multiuser OTFS on the Uplink},'' in \emph{Proc. IEEE 91st Veh. Tech. Conf.}, Antwerp, Belgium, 2020, pp. 1--5.

\bibitem{Ravi_Radar}
P.~Raviteja, K.~Phan, Y.~Hong, and E.~Viterbo, ``{Orthogonal Time Frequency Space (OTFS) Modulation Based Radar System},'' in \emph{2019 IEEE Radar Conference (RadarConf)}, Boston, MA, USA, 2019, pp. 1--6.

\bibitem{Liu_uplink_MIMO_ch_est_2020}
Y.~Liu, S.~Zhang, F.~Gao, J.~Ma, and X.~Wang, ``{Uplink-Aided High Mobility Downlink Channel Estimation Over Massive MIMO-OTFS System},'' \emph{IEEE Journal Sel. Areas Commun.}, vol.~38, no.~9, pp. 1994--2009, 2020.

\bibitem{Zhao_SBL_2020}
L.~Zhao, W.-J. Gao, and W.~Guo, ``{Sparse Bayesian Learning of Delay-Doppler Channel for OTFS System},'' \emph{IEEE Commun. Lett.}, vol.~24, no.~12, pp. 2766--2769, 2020.

\bibitem{Shi_TWC_OTFS}
D.~Shi, W.~Wang, L.~You, X.~Song, Y.~Hong, X.~Gao, and G.~Fettweis, ``{Deterministic Pilot Design and Channel Estimation for Downlink Massive MIMO-OTFS Systems in Presence of the Fractional Doppler},'' \emph{IEEE Trans. Wireless Commun.}, vol.~20, no.~11, pp. 7151--7165, 2021.

\bibitem{Suraj_BSBL_2021}
S.~Srivastava, R.~K. Singh, A.~K. Jagannatham, and L.~Hanzo, ``{Bayesian Learning Aided Sparse Channel Estimation for Orthogonal Time Frequency Space Modulated Systems},'' \emph{IEEE Trans. Veh. Tech.}, vol.~70, no.~8, pp. 8343--8348, 2021.

\bibitem{learning_choks}
S.~R. Mattu and A.~Chockalingam, ``{Learning based Delay-Doppler Channel Estimation with Interleaved Pilots in OTFS},'' in \emph{2022 IEEE 96th Vehicular Technology Conference (VTC2022-Fall)}, 2022, pp. 1--6.

\bibitem{naikoti2021signal_choks}
A.~Naikoti and A.~Chockalingam, ``{Signal detection and channel estimation in OTFS},'' \emph{ZTE Communications}, vol.~19, no.~4, pp. 16--33, 2021.

\bibitem{Mishra_superimposed_pilots_ch_est_2022}
H.~B. Mishra, P.~Singh, A.~K. Prasad, and R.~Budhiraja, ``{OTFS Channel Estimation and Data Detection Designs With Superimposed Pilots},'' \emph{IEEE Trans Wireless Commun.}, vol.~21, no.~4, pp. 2258--2274, 2022.

\bibitem{jesbin_superimposed}
F.~Jesbin, S.~Rao~Mattu, and A.~Chockalingam, ``{Sparse Superimposed Pilot Based Channel Estimation in OTFS Systems},'' in \emph{2023 IEEE Wireless Communications and Networking Conference (WCNC)}, 2023, pp. 1--6.

\bibitem{tharaj_OTSM}
T.~Thaj, E.~Viterbo, and Y.~Hong, ``{Orthogonal Time Sequency Multiplexing Modulation: Analysis and Low-Complexity Receiver Design},'' \emph{IEEE Transactions on Wireless Communications}, vol.~20, no.~12, pp. 7842--7855, 2021.

\bibitem{Thomas_ChaEst}
A.~Thomas, K.~Deka, P.~Raviteja, and S.~Sharma, ``{Convolutional Sparse Coding Based Channel Estimation for OTFS-SCMA in Uplink},'' \emph{IEEE Transactions on Communications}, vol.~70, no.~8, pp. 5241--5257, 2022.

\bibitem{Li_Detection}
L.~Li, H.~Wei, Y.~Huang, Y.~Yao, W.~Ling, G.~Chen, P.~Li, and Y.~Cai, ``{ simple two-stage equalizer with simplified orthogonal time frequency space modulation over rapidly time-varying channels},'' in \emph{available in arXiv}, 2017.

\bibitem{Zemen_Detection}
T.~Zemen, M.~Hofer, and D.~Loeschenbrand, ``{Low-complexity equalization for orthogonal time and frequency signaling (OTFS)},'' in \emph{available in arXiv}, 2017.

\bibitem{raviteja2018interference}
P.~Raviteja, K.~T. Phan, Y.~Hong, and E.~Viterbo, ``{Interference Cancellation and Iterative Detection for Orthogonal Time Frequency Space Modulation},'' \emph{IEEE Trans. Wireless Commun.}, vol.~17, no.~10, pp. 6501--6515, 2018.

\bibitem{Tiwari_Detection}
S.~Tiwari, S.~Das, and V.~Rangamgar, ``{Low complexity LMMSE receiver for OTFS},'' \emph{IEEE Commun. Lett.}, vol.~23, no.~12, p. 2205–2209, 2019.

\bibitem{surabhi2019low_comp_lin_equal}
G.~D. Surabhi and A.~Chockalingam, ``{Low-Complexity Linear Equalization for OTFS Modulation},'' \emph{IEEE Commun. Lett.}, vol.~24, no.~2, pp. 330--334, 2020.

\bibitem{pandey2021low}
B.~C. Pandey, S.~K. Mohammed, P.~Raviteja, Y.~Hong, and E.~Viterbo, ``{Low Complexity Precoding and Detection in Multi-User Massive MIMO OTFS Downlink},'' \emph{IEEE Trans. Veh. Tech.}, vol.~70, no.~5, pp. 4389--4405, 2021.

\bibitem{Zou_2021_ICC}
T.~Zou, W.~Xu, H.~Gao, Z.~Bie, Z.~Feng, and Z.~Ding, ``{Low-complexity linear equalization for OTFS systems with rectangular waveforms},'' in \emph{Proc. IEEE Int. Conf. Commun. Workshops (ICC Workshops)}, Montreal, QC, Canada, 2021, pp. 1--6.

\bibitem{ge2021receiveroversampling}
Y.~Ge, Q.~Deng, P.~C. Ching, and Z.~Ding, ``{Receiver Design for OTFS with a Fractionally Spaced Sampling Approach},'' \emph{IEEE Trans. Wireless Commun.}, vol.~20, no.~7, pp. 4072--4086, 2021.

\bibitem{Hanzo_TVT}
L.~Xiang, Y.~Liu, L.-L. Yang, and L.~Hanzo, ``{Gaussian Approximate Message Passing Detection of Orthogonal Time Frequency Space Modulation},'' \emph{IEEE Trans. Veh. Tech.}, vol.~70, no.~10, pp. 10\,999--11\,004, 2021.

\bibitem{Zhang_2021_AMP}
Y.~Zhang, Q.~Zhang, L.~Zhang, C.~He, and X.~Tian, ``{A Low-Complexity Approximate Message Passing Equalizer for OTFS system},'' in \emph{Proc. IEEE/CIC Int. Conf. Commun. China}, Xiamen, China, 2021, pp. 449--454.

\bibitem{Li_2021_EP}
H.~Li, Y.~Dong, C.~Gong, Z.~Zhang, X.~Wang, and X.~Dai, ``{Low Complexity Receiver via Expectation Propagation for OTFS Modulation},'' \emph{IEEE Commun. Lett.}, vol.~25, no.~10, pp. 3180--3184, 2021.

\bibitem{Yuan_2022_UAMP}
Z.~Yuan, F.~Liu, W.~Yuan, Q.~Guo, Z.~Wang, and J.~Yuan, ``{Iterative Detection for Orthogonal Time Frequency Space Modulation With Unitary Approximate Message Passing},'' \emph{IEEE Trans. on Wireless Commun.}, vol.~21, no.~2, pp. 714--725, 2022.

\bibitem{tharaj_rake_MRC_conf}
T.~Thaj and E.~Viterbo, ``{Low Complexity Iterative Rake Detector for Orthogonal Time Frequency Space Modulation},'' in \emph{Proc. IEEE Wireless Commun. Networking Conf.}, Seoul, Korea (South), 2020, pp. 1--6.

\bibitem{tharaj_rake_MRC_journal}
------, ``{Low Complexity Iterative Rake Decision Feedback Equalizer for Zero-Padded OTFS Systems},'' \emph{IEEE Trans. Veh. Tech.}, vol.~69, no.~12, pp. 15\,606--15\,622, 2020.

\bibitem{tharaj_universal_MRC}
T.~Thaj, E.~Viterbo, and Y.~Hong, ``{General I/O Relations and Low-Complexity Universal MRC Detection for All OTFS Variants},'' \emph{IEEE Access}, vol.~10, pp. 96\,026--96\,037, 2022.

\bibitem{PP_Oversam_MRC}
P.~Priya, E.~Viterbo, and Y.~Hong, ``{Low Complexity MRC Detection for OTFS Receiver with Oversampling},'' \emph{IEEE Transactions on Wireless Communications}, pp. 1--1, 2023.

\bibitem{DNN_Cholk}
A.~Naikoti and A.~Chockalingam, ``{Low-complexity Delay-Doppler Symbol DNN for OTFS Signal Detection},'' in \emph{Proc. IEEE 93rd Veh. Tech. Conf.}, Helsinki, Finland, 2021, pp. 1--6.

\bibitem{CNN_Enku}
Y.~K. Enku, B.~Bai, F.~Wan, C.~Guyo, I.~N. Tiba, C.~Zhang, and S.~Li, ``{Two-dimensional convolutional neural network-based signal detection for OTFS systems},'' \emph{IEEE Wireless Commun. Lett.}, vol.~10, no.~11, p. 2514–2518, 2021.

\bibitem{Calderbank_learning}
Z.~Zhou, L.~J. Liu, J.~Xu, and R.~Calderbank, ``{Learning to Equalize OTFS},'' \emph{IEEE Transactions on Wireless Communications}, vol.~21, no.~9, pp. 7723 -- 7736, 2022.

\bibitem{Fan_survey}
Z.~Zhang, L.~Heng, W.~Qianli, and P.~Fan, ``{A Survey on Low Complexity Detectors for OTFS SystemsDetectors for OTFS Systems},'' \emph{ZTE Communications}, vol.~4, no.~19, pp. 3--15, 2021.

\bibitem{dual_chirp}
X.~Geng, Z.~Liu, and H.~Wu, ``{Timing synchronization based on Radon--Wigner transform of chirp signals for OTFS systems},'' \emph{Physical Communication}, p. 102161, 2023.

\bibitem{Saif_IDZT}
S.~K. Mohammed, ``Time-domain to delay-doppler domain conversion of otfs signals in very high mobility scenarios,'' \emph{IEEE Transactions on Vehicular Technology}, vol.~70, no.~6, pp. 6178--6183, 2021.

\bibitem{ETU}
``{LTE; Evolved Universal Terrestrial Radio Access (E-UTRA); Base Station (BS) radio transmission and reception},'' \emph{3GPP TS 36.104 version 14.3.0 Release 14}, 2017.

\end{thebibliography}
 \begin{IEEEbiography}[{\includegraphics[width=1in,height=1.25in,clip,keepaspectratio]{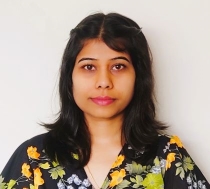}}]{Preety Priya} (GS'18--M'22) is currently an Assistant Professor at the Department of Electronics and Communication Eng., National Institute of Technology Calicut, India. She received her M.Tech. degree in telecommunication systems engineering in 2014, and Ph.D. degree in 2022 from the Indian Institute of Technology Kharagpur. From 2022 to 2024 she was a Research Fellow with the Department of Electrical and Computer Systems Engineering, Monash University, Melbourne, Australia. From 2014 to 2016, she worked as an Assistant Professor with Kalinga Institute of Industrial Technology, Bhubaneswar, India.  She was awarded the Qualcomm Innovation Fellowship from Qualcomm in 2017. Her research interests include applications of 5G mmWave MIMO wireless technology, sparse signal processing, nonlinear estimation, and OTFS systems. 
\end{IEEEbiography}

\begin{IEEEbiography}[{\includegraphics[width=1in,height=1.2in,clip]{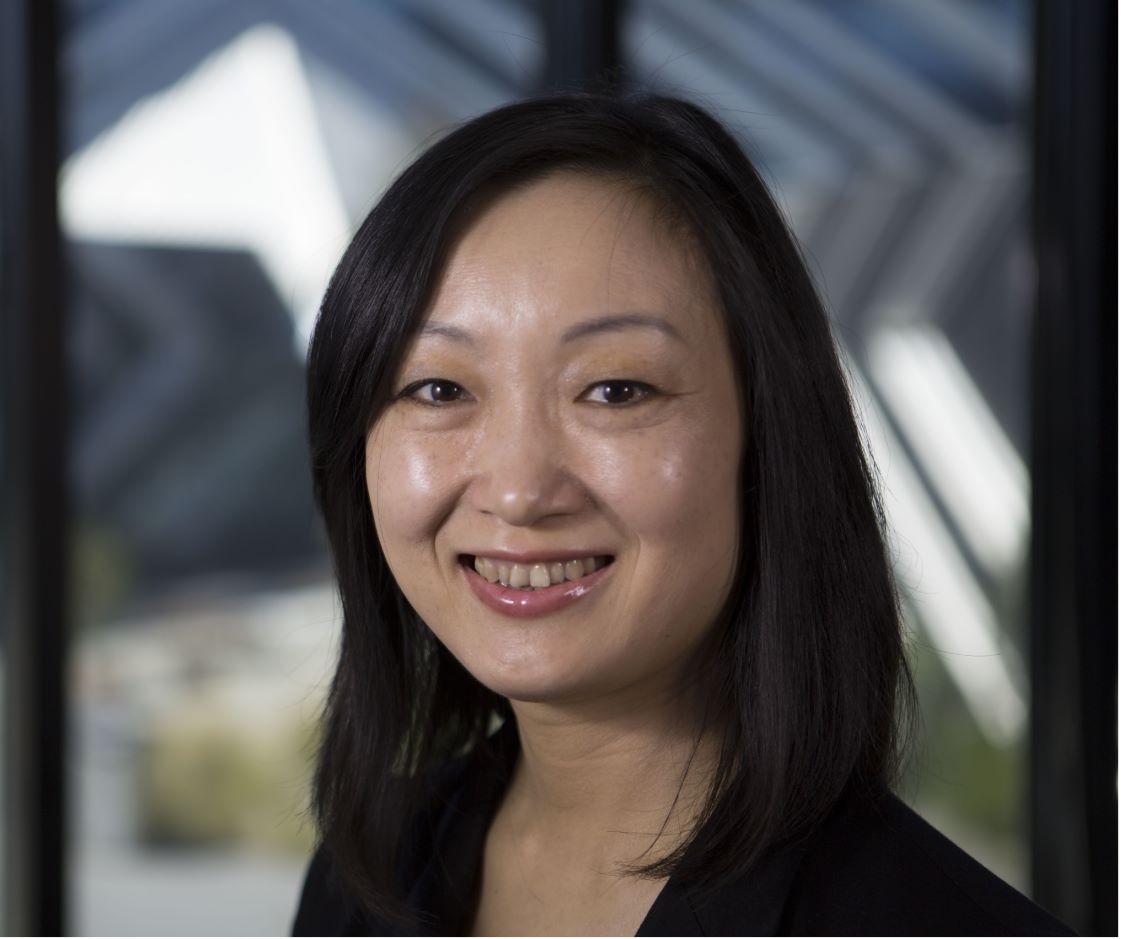}}]{Yi Hong}(S'00--M'05--SM'10)
is currently an associate professor at the Department of Electrical and Computer Systems Eng.,
Monash University, Melbourne, Australia.
She obtained her Ph.D. degree in Electrical Engineering and Telecommunications 
from the University of New South Wales (UNSW), Sydney, and received   
the {\em NICTA-ACoRN Earlier Career Researcher Award} at the {\em Australian Communication
Theory Workshop}, Adelaide, Australia, 2007. She served on the Australian Research Council College of Experts (2018-2020).
 
Prof Hong is currently an Associate Editor (AE) for the {\em IEEE Transactions on Green Communications and Networking}, 
and was the AE for the {\em IEEE Wireless Communication Letters} and {\em Transactions on Emerging Telecommunications Technologies (ETT)}.
She served as the Tutorial Chair of {\em IEEE International Symposium on Information Theory}, Melbourne, 2021,
and the General Co-Chair for the {\em OTFS workshops} in the {\em 2019-22 IEEE Communications Conferences},  
the {\em OTFS workshop} in the {\em IEEE Vehicular Technology Conference-Fall}, 2021, and the {\em IEEE Information Theory Workshop}, Hobart, 2014.
Her research interests include communication theory, coding and information theory with applications to telecommunication engineering.
\end{IEEEbiography}
\begin{IEEEbiography}[{\includegraphics[width=1in,height=1.25in,clip,keepaspectratio]{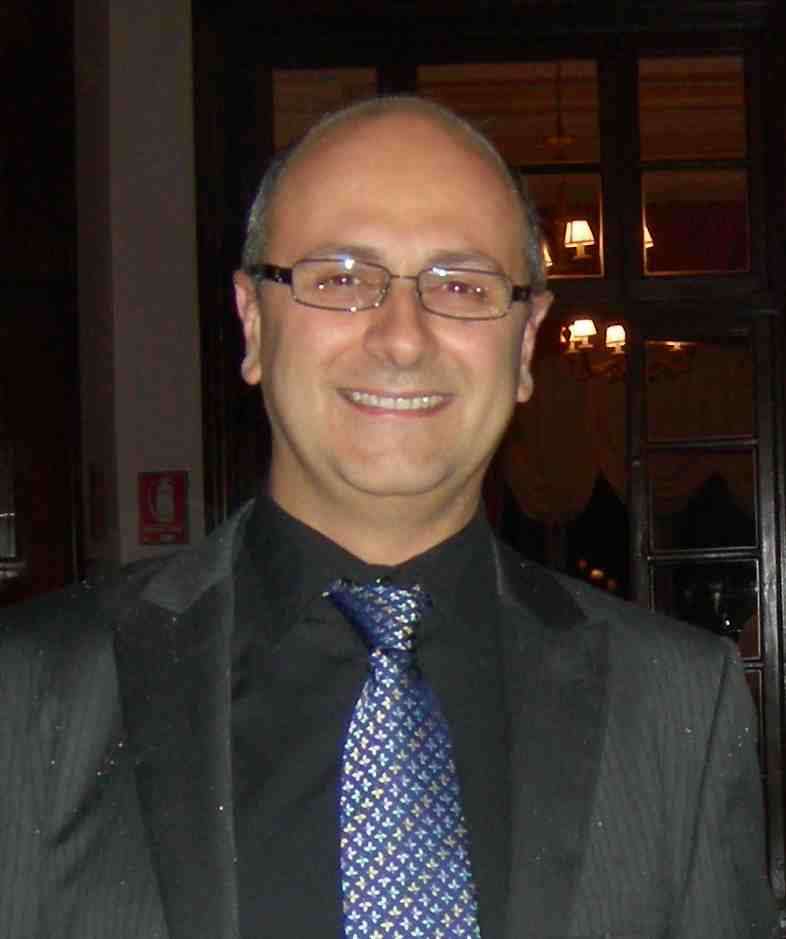}}]{Emanuele Viterbo}(M'95--SM'04--F'11)
is currently a professor in the ECSE Department and an Associate Dean in Graduate Research at
Monash University, Melbourne, Australia.
He received his Ph.D. in 1995 in Electrical Engineering, from the Politecnico di Torino, Torino, Italy.
From 1990 to 1992 he was with the European Patent Office, The Hague, The Netherlands, as a patent examiner
in the field of dynamic recording and error-control coding. Between 1995 and 1997 he held a post-doctoral
position in the Dipartimento di Elettronica of the Politecnico di Torino. In 1997-98 he was a post-doctoral research fellow
in the Information Sciences Research Center of AT T Research, Florham Park, NJ, USA.
From 1998-2005, he worked as Assistant Professor and then Associate Professor, in Dipartimento di Elettronica at Politecnico di Torino.
From 2006-2009, he worked in DEIS at University of Calabria, Italy, as a Full Professor.
Prof. Emanuele Viterbo is an ISI Highly Cited Researcher since 2009. He is an Associate Editor of  \emph{IEEE Trans. on Information Theory}, \emph{European Transactions on Telecommunications, and Journal of Communications and Networks}, and a Guest Editor of \emph{IEEE J. of Selected Topics in Signal Processing: Special Issue Managing Complexity in Multiuser MIMO Systems}.
Prof. Emanuele Viterbo was awarded a NATO Advanced Fellowship in 1997 from the Italian National Research Council.
His main research interests are in lattice codes for the Gaussian and fading channels, algebraic coding theory,
algebraic space-time coding, digital terrestrial television broadcasting, digital magnetic recording, and irregular sampling.
\end{IEEEbiography}

\end{document}